\documentclass[aps,prb,twocolumn,superscriptaddress,showpacs,amsmath,amssymb]{revtex4-1}

\usepackage{amssymb,amsbsy,amsmath}
\usepackage{graphicx}
\usepackage{dcolumn}
\usepackage{bm}
\usepackage{subfigure}
\usepackage{color}
\usepackage{textcomp}
\usepackage[colorlinks,bookmarks=false,citecolor=blue,linkcolor=red,urlcolor=blue]{hyperref}

\newcommand{\op}[1]{%
    \fontdimen12\textfont3=2pt\fontdimen12\scriptfont3=1.4pt%
    \!\null\mathop{\vphantom{#1}\smash{#1}}\limits_{\sim}\null\!}
\newcommand{\xref}[1]{\protect\ref{#1}}
\newcommand{\figref}[1]{Fig.~\protect\ref{#1}}
\newcommand{\fmref}[1]{(\protect\ref{#1})}

\begin{document}
\title{Influence of intermolecular interactions on
  magnetic observables}

\author{J\"urgen Schnack}
\email{jschnack@uni-bielefeld.de}
\affiliation{Fakult\"at f\"ur Physik, Universit\"at Bielefeld, Postfach 100131, D-33501 Bielefeld, Germany}

\date{\today}

\begin{abstract}
Very often it is an implied paradigm of molecular magnetism that
magnetic molecules in a crystal interact so weakly that
measurements of dc magnetic observables reflect ensemble
properties of single molecules. But the number of cases where
the assumption of virtually non-interacting molecules does not
hold grows steadily. A deviation from the non-interacting case 
can especially clearly be seen in clusters with
antiferromagnetic couplings, where steps of the low-temperature
magnetization 
curve are smeared out with increasing intermolecular
interaction. In this investigation we demonstrate with examples
in one-, 
two, and three space dimensions how intermolecular interactions
influence typical magnetic observables such as magnetization,
susceptibility and specific heat.
\end{abstract}


\maketitle

\section{Introduction}
\label{sec-1}

The statement, that the assumption of non-interacting magnetic
molecules constitutes a paradigm, is of course exaggerated. In
the more specific scientific community which deals with magnetic
molecules for quantum computing the question of intermolecular
(and other unwanted) interactions is of course of utmost
importance.\cite{ARM:PRL07,CLT:PRL10,CGT:PRA14,PCM:NC14,LGB:PRL14,LKC:JPSJ14,YNS:PCCP15}
Also in low-dimensional magnetism 
the question arises for instance in connection with dimerization
or interchain as well as interlattice
interactions.\cite{Ogu:PR64,TaY:JPSJ70,TYM:JPSJ70,OgB:JPSJ81,NTS:PRB92,KTH:Science02,SSS:PRB03,SRI:PRB05,YTH:PRL05,LWL:JACS07,NDF:DT08,BLT:ICA08,MMY:JACS09,AVA:PRB11}
Intermolecular exchange is also carefully taken care of when
designing magnetic multi-qubit devices. For these cases the
intermolecular interactions are even made
switchable.\cite{MMY:JACS09}

In this paper we would like to approach the problem from a
somewhat different perspective. We would like to ask how big
intermolecular interactions have to be in order to modify dc
magnetic observables so drastically that the fingerprint of the
underlying molecular subunits is masked. We concentrate
our investigations on molecules with an antiferromagnetic
intramolecular coupling. For these cases the low-temperature
magnetization curves consist of clearly spaced
steps,\cite{ShB:JAP02} which disappear with increasing
antiferromagnetic intermolecular interaction. Such a behavior
was observed in several recent investigations, compare
e.g. Refs.~\onlinecite{SPK:PRB08,KKA:CC15,PPS:CPC15}, 
and interpreted in various ways. We don't offer a solution to
specific problems, but we would like to present
order-of-magnitude calculations that show for which ratios of
inter- and intramolecular interactions magnetization steps of
finite clusters disappear. We demonstrate that the space
dimension of the embedding plays a strong role.

The evaluation of magnetic properties of interacting magnetic
molecules constitutes a massive quantum many-body problem, even
if only spin Hamiltonians are considered. That's why only two to three
dozens of spins $s=1/2$ can be modeled numerically exactly by
diagonalizing a spin Hamiltonian even when symmetries are
employed.\cite{DGP:IC93,GaP:GCI93,BCC:IC99,BeG:EPR,Tsu:group_theory,Wal:PRB00,Tsu:ICA08,BBO:PRB07,ScS:PRB09,ScS:IRPC10}
But for
non-frustrated spin systems quantum Monte
Carlo\cite{SaK:PRB91,San:PRB99,SyS:PRE02} (QMC) provides quasi
exact thermodynamic observables. We therefore restrict our
investigations to a series of such systems, which should be
sufficient for the purpose of this paper. The clear advantage is
that we do not rely on mean field
approximations.\cite{TaY:JPSJ70,OgB:JPSJ81,Sch:PRL96} 

The result of our investigations is that single molecule
signatures are washed out in the magnetization if the
intermolecular interactions are stronger than about 10~\% of the
intramolecular interactions. The specific value depends on the
space dimension of the embedding, e.g. for intermolecular interactions
in three dimensions a smaller intermolecular interaction is
needed to mask the molecular behavior than in lower dimensions.
We compare some of our results with the scenario of J-strain,
that sometimes is also taken into consideration for the
interpretation of experimental data.

Finally, we would like to draw the readers attention to related
investigations. An important related problem is given by the 
influence of interchain or interplane interactions on magnetic
observables and in particular ordering temperatures in
antiferromagnetic systems. Such questions are also dealt with
by means of QMC for instance in
Refs.~\onlinecite{SSS:PRB03,YTH:PRL05,BLT:ICA08}. 
Random-exchange quantum Heisenberg antiferromagnets on a square
lattice have been investigated by QMC in Ref.~\onlinecite{LWL:PRB06}.
The influence
of the embedding medium on ground state properties of a
Heisenberg star system was investigated in
Ref.~\onlinecite{RVK:JPAMG96}.

The article is organized as follows. In Section~\xref{sec-2} the
theoretical framework is explained. Section~\xref{sec-3} 
presents magnetization, susceptibility and specific heat for
dimers, squares and cubes embedded in one, two and three
dimensions, respectively, whereas Section~\xref{sec-4} discusses
the magnetism of dimers with intermolecular interactions in one,
two and three dimensions. 
In Section~\xref{sec-5} a comparison with the scenario of
J-strain is presented. The article closes with summary and
outlook.

\section{Hamiltonian and calculational scheme}
\label{sec-2}

The investigated spin systems are modeled by a Heisenberg
Hamiltonian augmented with a Zeeman term, i. e.  
\begin{eqnarray}
\label{E-2-1}
\op{H}
&=&
-
\sum_{i<j}\;
{J}_{ij}
\op{\vec{s}}_i \cdot \op{\vec{s}}_j
+
g \mu_B\, B\,
\sum_{i}\;
\op{s}^z_i
\ .
\end{eqnarray}
Thermodynamic observables are evaluated by means of 
quantum Monte Carlo\cite{SaK:PRB91,San:PRB99,SyS:PRE02} (QMC)
using the ALPS package.\cite{ALPS:JMMM07,BCE:JSMTE11} The
prefactors in \fmref{E-2-1} are chosen according to the 
convention used in ALPS, in particular a negative
Heisenberg exchange corresponds to an antiferromagnetic
interaction. Without loss of generality 
$s=1/2$ is chosen as spin quantum number, and the spectroscopic 
splitting factor is taken as $g=2$ for all spins. 

In our QMC calculations we choose $N=100$ for one-dimensional
problems, $N=10\times 10$ for two-dimensional problems as well
as $N=10\times 10\times 10$ for three-dimensional problems. In
all cases periodic boundary conditions are applied. For
magnetization curves we use 100000 steps for thermalization and
another 100000 steps for sampling. In case of susceptibility and
specific heat functions of temperature at $B=0$ thermalization
is done with 10000000 steps and sampling with 200000000
steps. Although we used $\epsilon=0.1$ for the latter functions,
convergence was very slow at the lowest temperatures, which can
be seen in the upcoming plots. Fortunately, this does not alter
our conclusions.

\section{Magnetic observables for one-, two-, and
  three-dimensional intermolecular interactions}
\label{sec-3}

\begin{figure}[ht!]
\centering
\includegraphics*[clip,width=60mm]{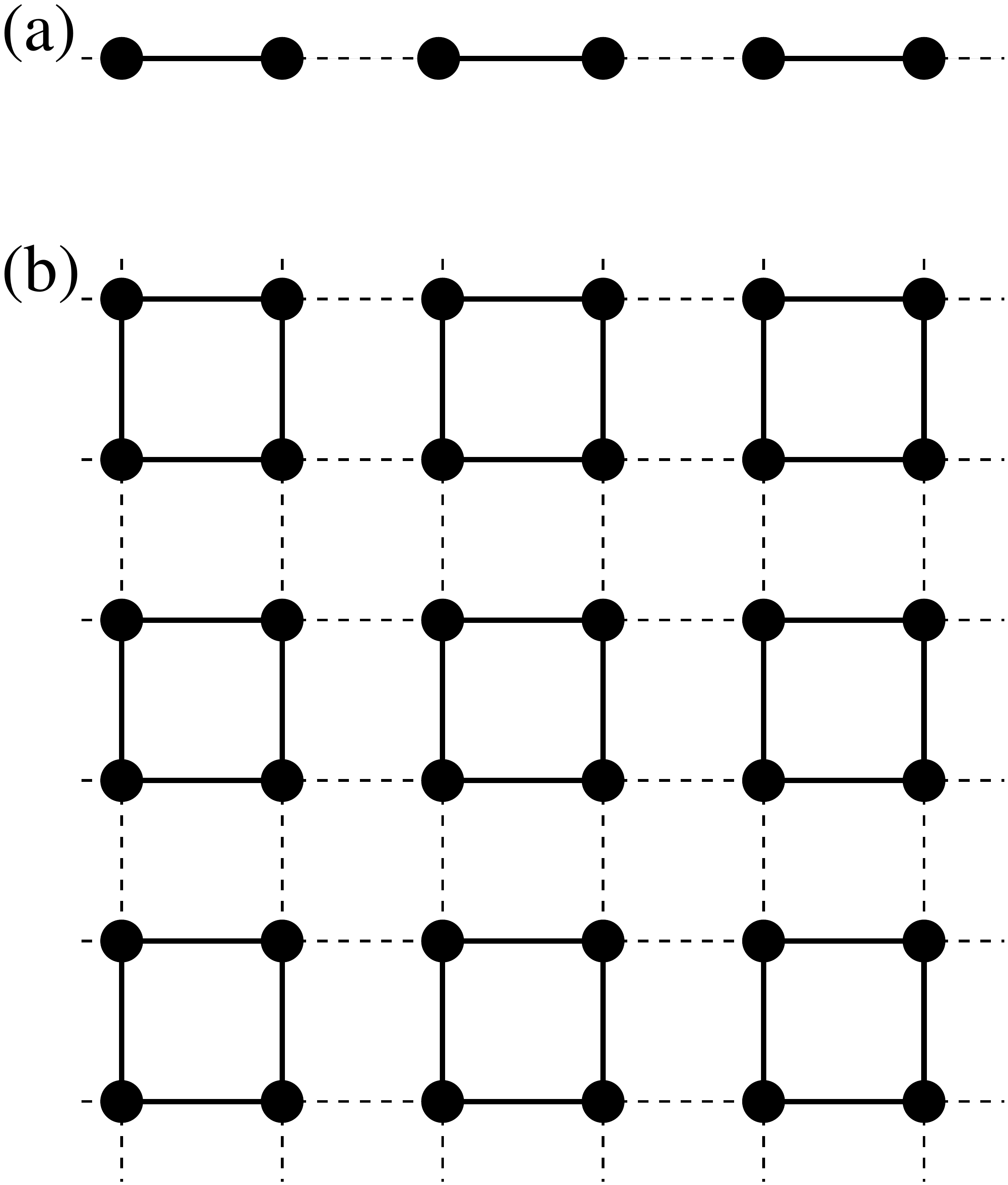}
\caption{(Color online) Schematic structure of the
  one-dimensional (a) and the two-dimensional bipartite lattice; solid bonds
  depict interactions $J_1$, dashed bonds $J_2$.}   
\label{intermolecular-f-1}
\end{figure}

In the following magnetic observables are presented for small
magnetic units (molecules) that are investigated in
dependence of the intermolecular interaction. These
investigations have been performed for couplings in one, two and
three space dimensions. Our main concern is the magnetization
curve, since this curve usually exhibits a very strong
dependence on whether and how subunits are magnetically
coupled.\cite{ShB:JAP02} For small systems the magnetization
curve exhibits clear fingerprints of the specific unit, such as
magnetization steps.  

We follow two rationals when comparing the behavior in different
space dimensions. In the first investigation each spin interacts
with the same number of intramolecular $J_1$ and intermolecular
$J_2$ bonds with its neighbors. This is realized by dimers in
one, squares in two and cubes in three dimensions. In addition,
the size of the singlet-triplet gap is the same for dimer and
square and almost the same (80~\%) for the cube. In a second
investigation the molecular unit is kept fixed as dimers, and
the dimension of the embedding is varied.

\subsection{One-dimensional system}
\label{sec-3-1}

One-dimensional systems have been very thoroughly investigated
over almost a century. The famous Bethe ansatz for the spin-1/2
Heisenberg chain and the resulting knowledge on observables as
well as the Haldane conjecture are cornerstones of this
research.\cite{Bet:ZP31,Hul:AMAFA38,Gri:PR64,Hal:PL83,Hal:PRL83,AfL:LMP86,AGS:JPA89,Klu:EPJB98}
Here we focus on the question how magnetic observables develop
with an increasing intermolecular interaction $J_2$ between
dimers that are coupled through $J_1$, compare
\figref{intermolecular-f-1}~(a). Both interactions are
antiferromagnetic.  Since we deal with four quantities, $J_1$,
$J_2$, $T$, and $B$, we
decided to assume some reasonable values throughout the article
that are common to 
materials in molecular magnetism, in particular we choose
$J_1=-10$~K.\footnote{The reader is of 
  course free to make all quantities dimensionless, e.g. by
  dividing by $J_1$.}
In some sense this investigation touches earlier works on
dimerized (or spin-Peierls) spin chains,\cite{Hid:PRB92,RAH:PRB96,UhS:PRB96,KnU:00} 
which investigate similar structures as
\figref{intermolecular-f-1}~(a) but usually for a fixed ratio of
interactions $J_2/J_1$ or some small interval of this ratio. 

\begin{figure}[ht!]
\centering
\includegraphics*[clip,width=60mm]{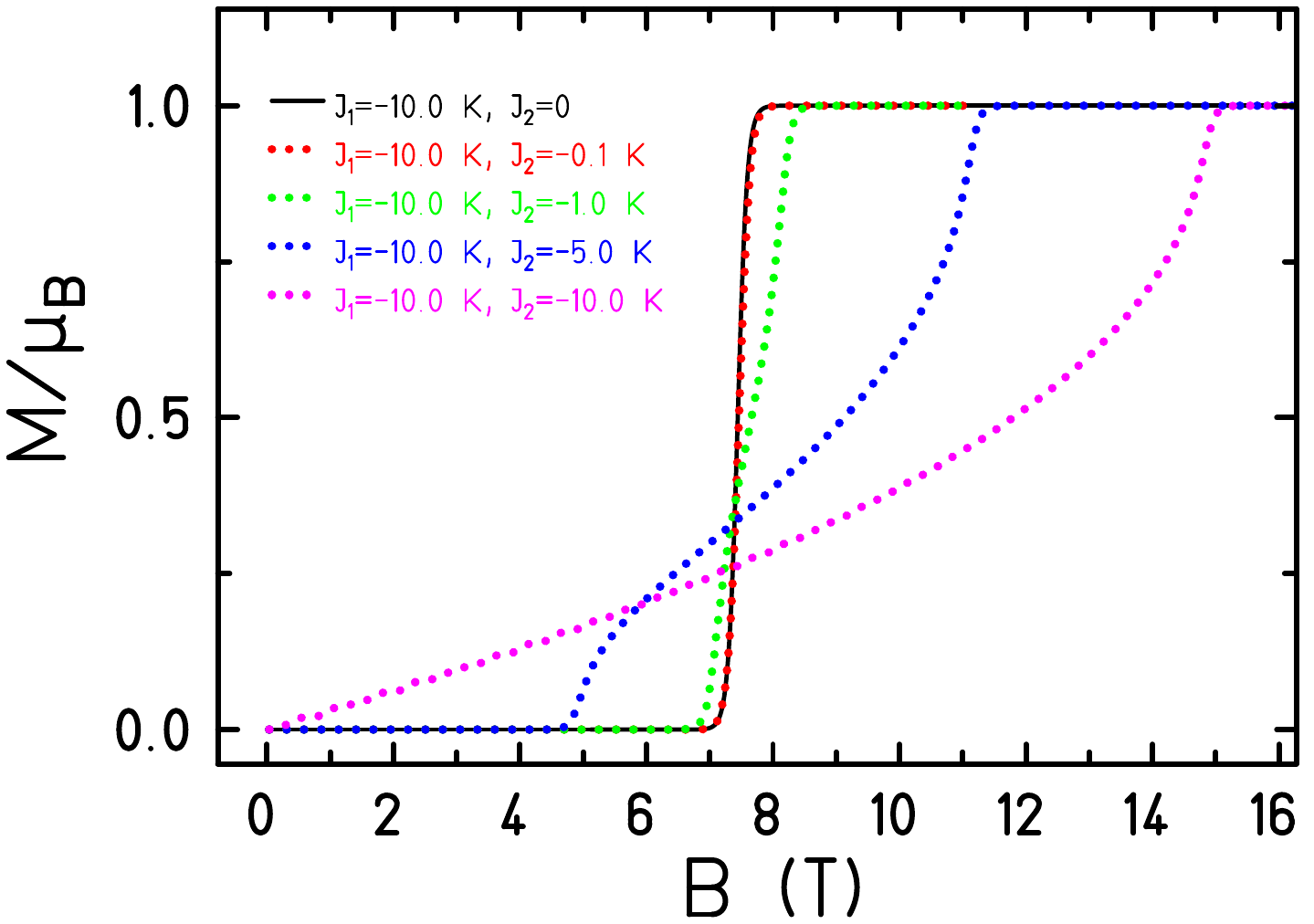}
\caption{(Color online) Low-temperature magnetization of the
  one-dimensional spin system shown in
  \figref{intermolecular-f-1}~(a) for various interdimer
  couplings $J_2$ and $T=0.1$~K.}   
\label{intermolecular-f-2}
\end{figure}

\begin{figure}[ht!]
\centering
\includegraphics*[clip,width=60mm]{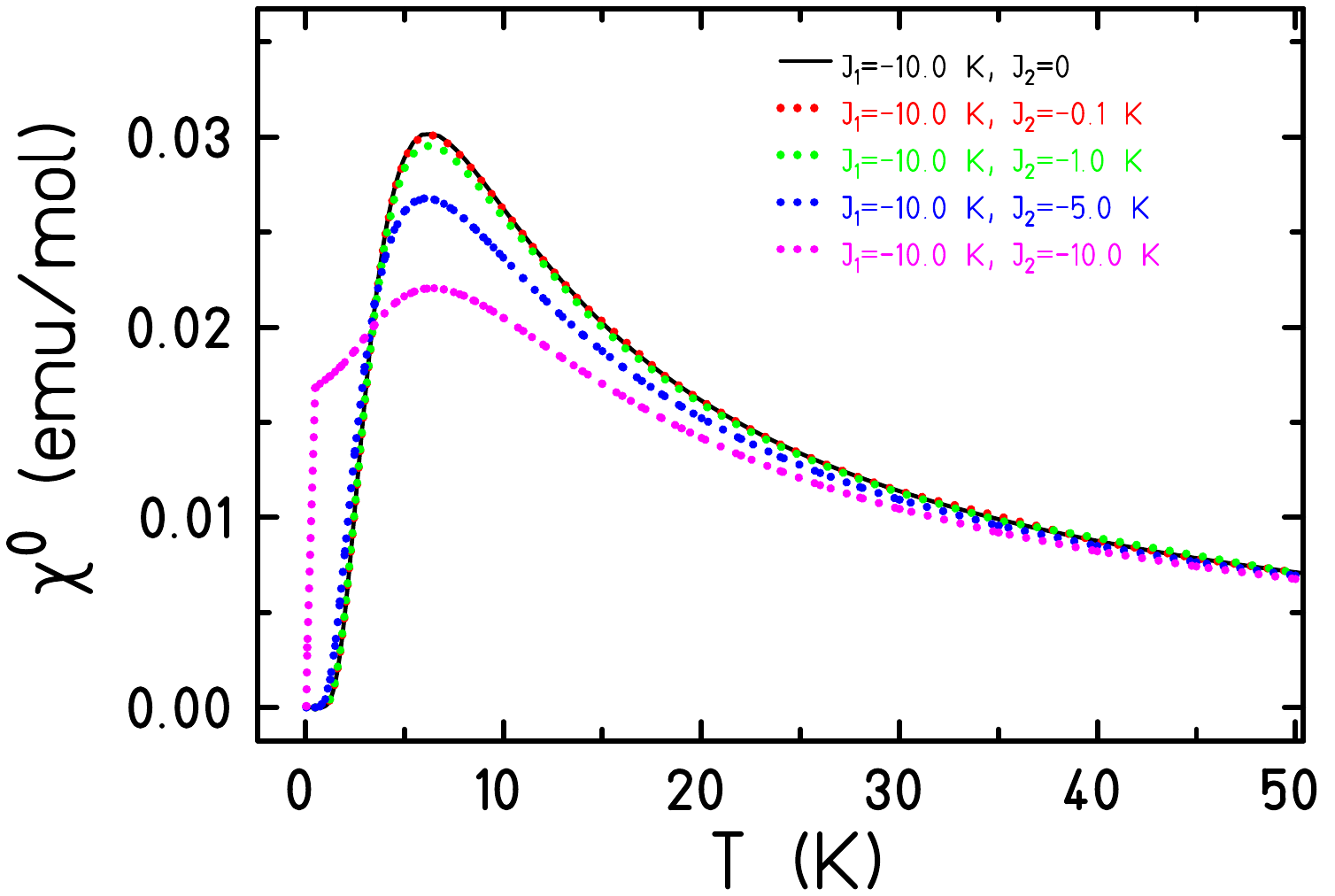}

\includegraphics*[clip,width=60mm]{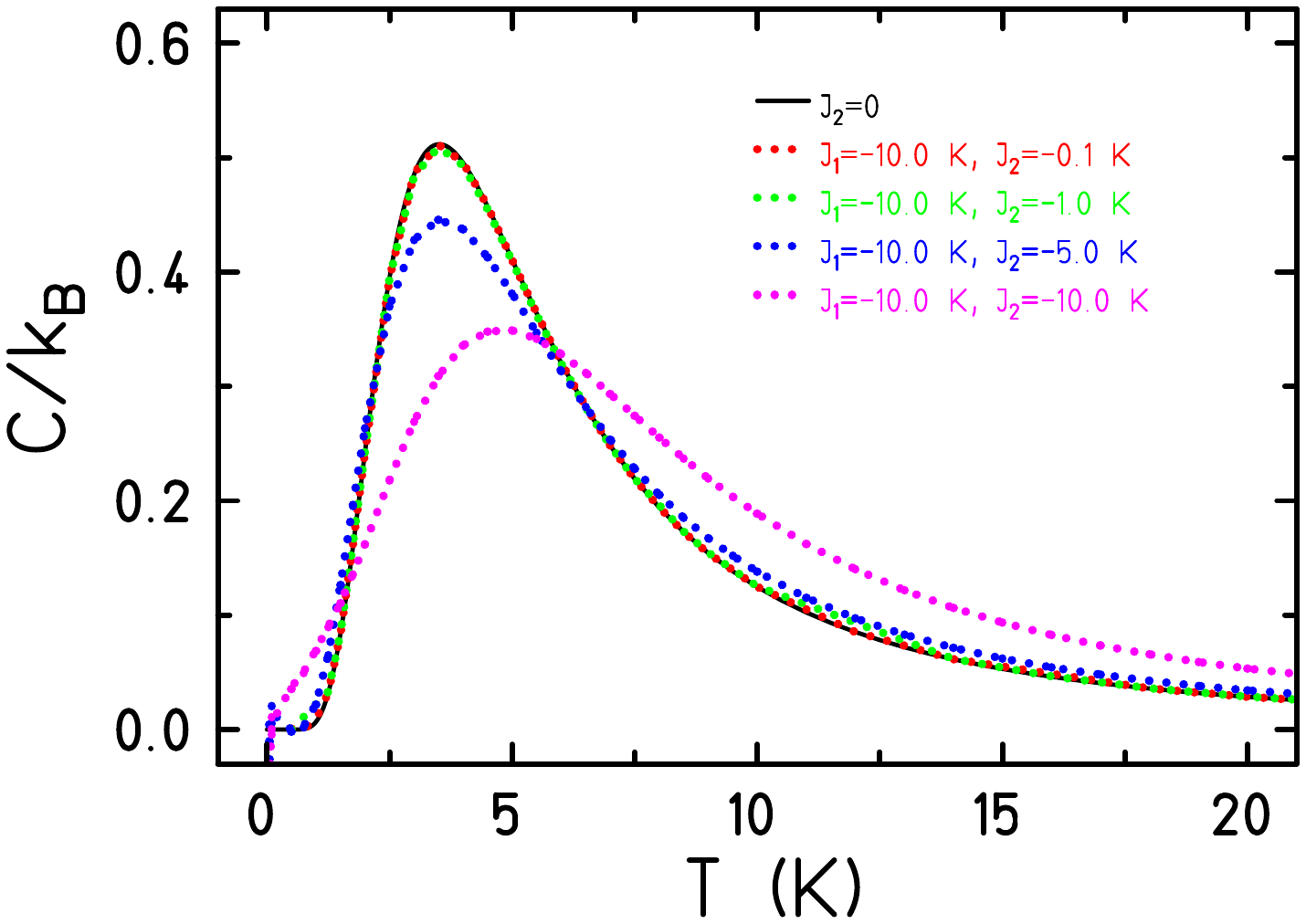}
\caption{(Color online) Zero-field susceptibility and specific
  heat of the 
  one-dimensional spin system shown in
  \figref{intermolecular-f-1}~(a) for various interdimer
  couplings $J_2$ and $B=0$. Fluctuations of $C$ at the lowest $T$
  result from very slow and thus insufficient convergence.}   
\label{intermolecular-f-3}
\end{figure}

For vanishing interdimer interaction $J_2=0$ the magnetization
curve at low temperatures features one jump to saturation at the
external field where the singlet and the lowest triplet level cross. 
This simple magnetization curve is characteristic for the af
dimer of two spins $s=1/2$. The jump is rather stable against an
increase of $J_2$; even at $J_2/J_1=0.5$ the curve still warps
around the former jump. For $J_2/J_1=1$ the limit of the
af Heisenberg chain is reached, which results in a gapless
continuous rise of the magnetization with increasing field. 

The robustness of the dimer properties is also reflected in the
susceptibility as well as specific heat functions, here in zero
field. Both functions do not change their dimer character up to
at least $J_2/J_1=0.5$. For $J_2/J_1=1$ they assume their
characteristic behavior known from the Bethe ansatz: the system
is gapless, and the specific heat shows Luttinger liquid
behavior, i.e. grows linearly with $T$ for low temperatures.

\subsection{Two-dimensional system}
\label{sec-3-2}

\begin{figure}[ht!]
\centering
\includegraphics*[clip,width=60mm]{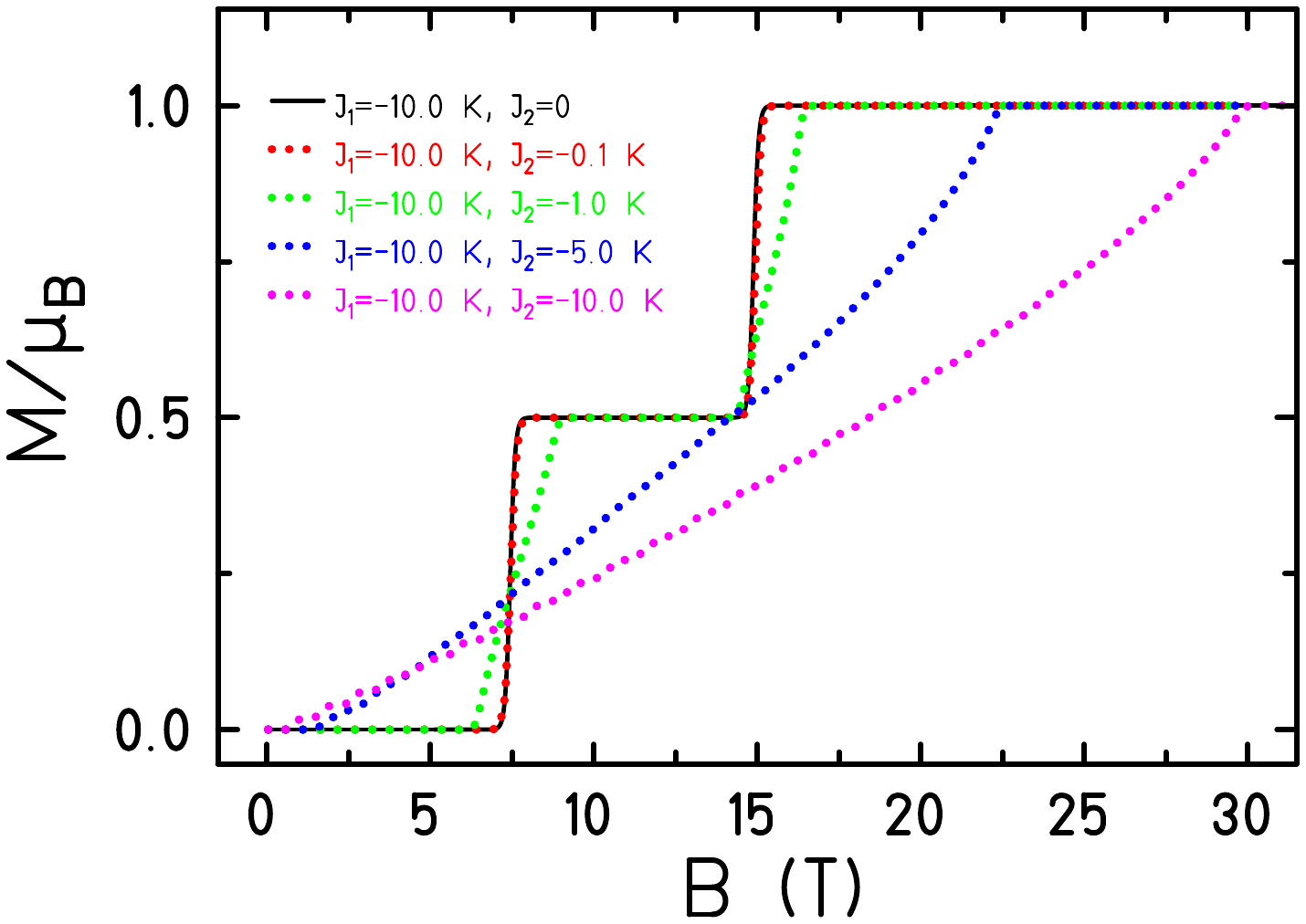}
\caption{(Color online) Low-temperature magnetization of the
  two-dimensional spin system shown in
  \figref{intermolecular-f-1}~(b) for various interdimer
  couplings $J_2$ and $T=0.1$~K.}   
\label{intermolecular-f-4}
\end{figure}

For the investigations in two dimensions we chose a square as
the unperturbed molecular system. The structure is shown in
\figref{intermolecular-f-1}~(b). A Heisenberg square with af
interaction shows two steps in the magnetization at fields $B_1$
and $B_2$ where the lowest triplet crosses the singlet and where
the lowest pentet crosses the lowest triplet. This is clearly
visible in \figref{intermolecular-f-4} (black solid curve).

With increasing ratio $J_2/J_1$ the step-like structure is more
quickly destroyed by the intermolecular interactions compared to
the one-dimensional case. For
$J_2/J_1=0.1$ the curve still warps around the former jumps, but
for $J_2/J_1=0.5$ it is already almost continuous and in its
character not much different from the magnetization curve of the
af square lattice.\cite{Man:RMP91,Ric:PRB93,San:PRB97} This
behavior is not so clearly reflected by 
the thermal functions. These are again rather stable against
variations of $J_2/J_1$, only the susceptibility displays the
change by a very different rise at low-temperature. The specific
heat does not display any feature since no ordering can occur in
one or two dimensions for non-zero temperature.\cite{MeW:PRL66}

\begin{figure}[ht!]
\centering
\includegraphics*[clip,width=60mm]{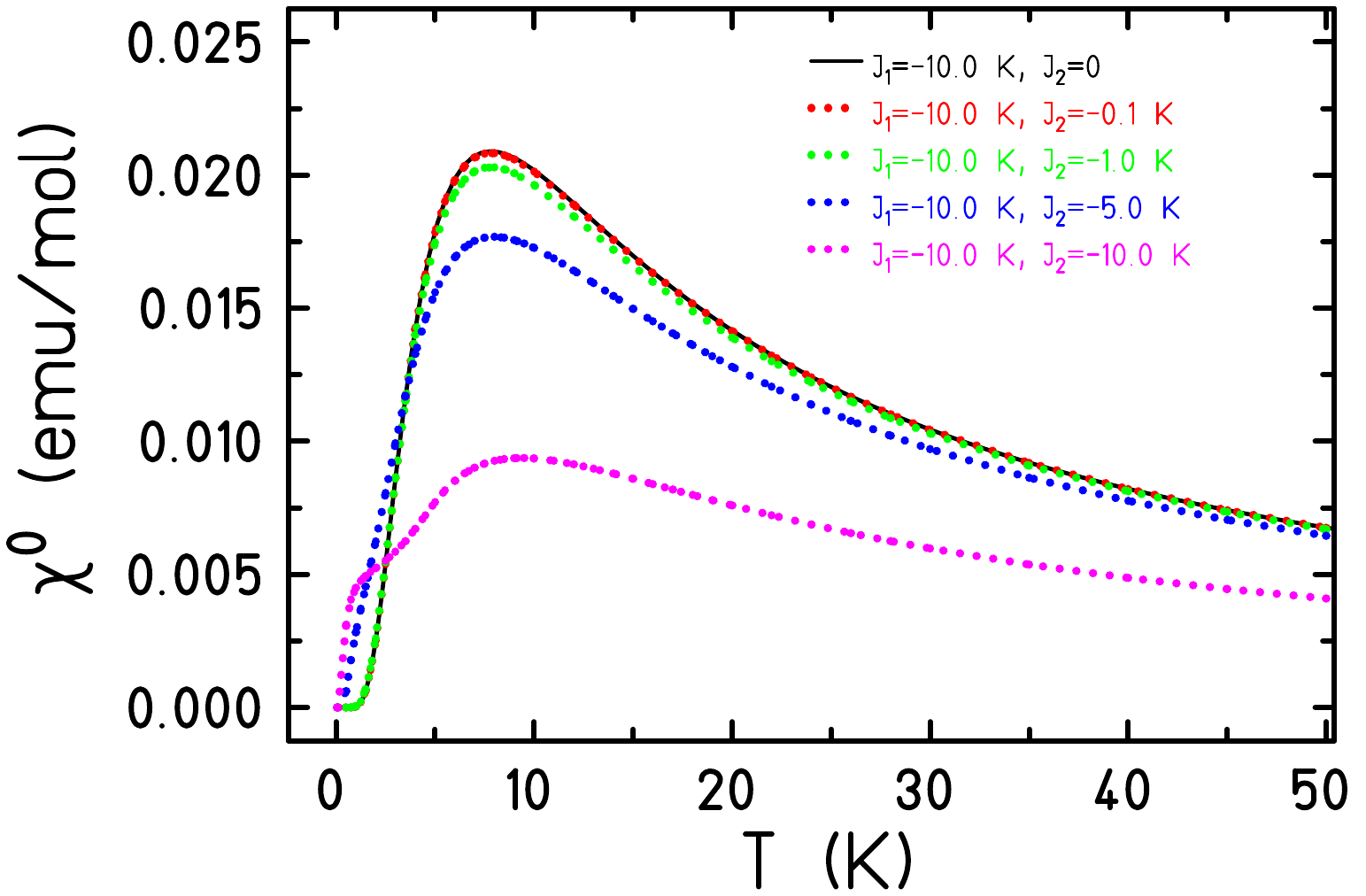}

\includegraphics*[clip,width=60mm]{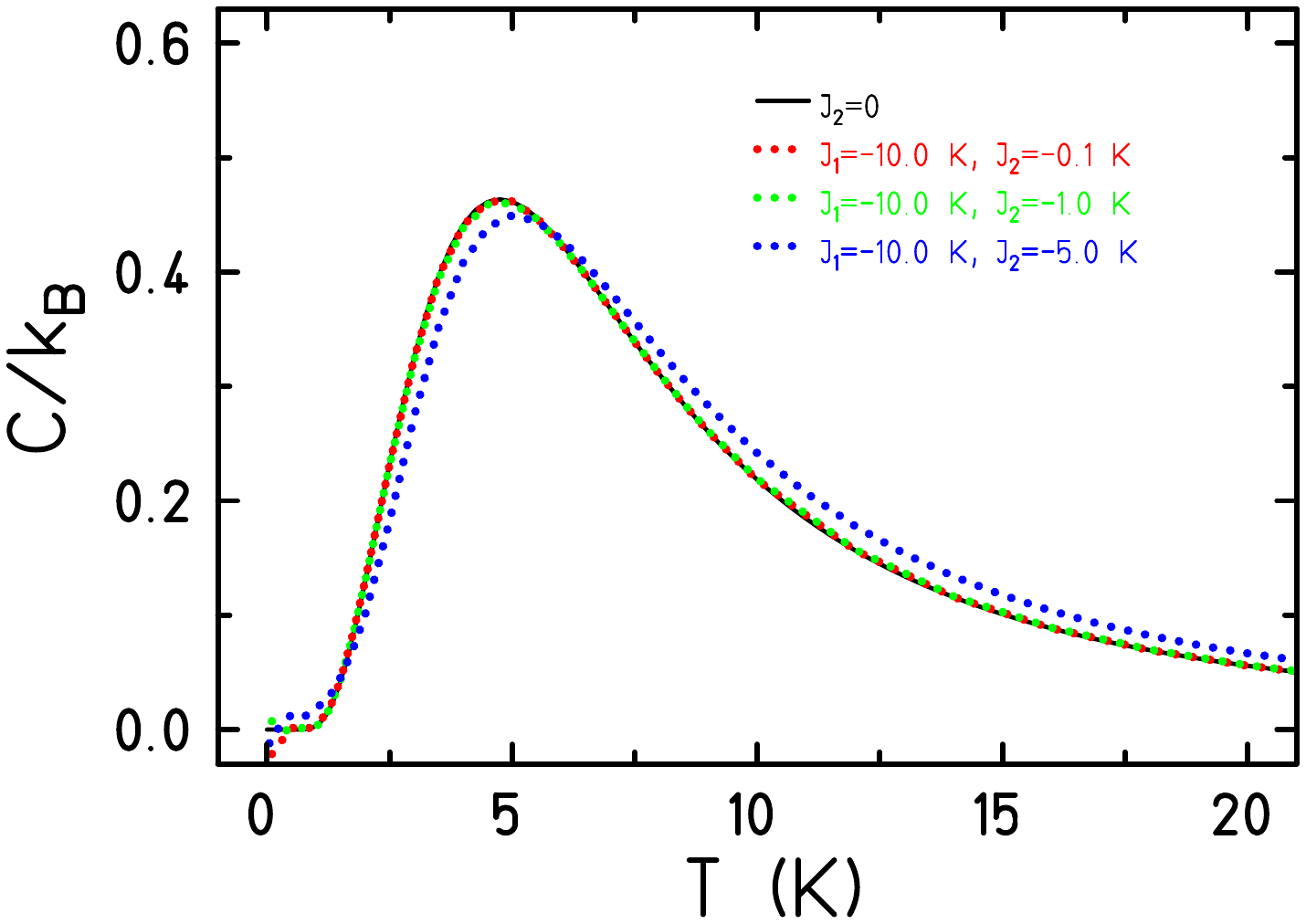}
\caption{(Color online) Zero-field susceptibility and specific heat of the 
  two-dimensional spin system shown in
  \figref{intermolecular-f-1}~(b) for various interdimer
  couplings $J_2$ and $B=0$.}   
\label{intermolecular-f-5}
\end{figure}

\subsection{Three-dimensional system}
\label{sec-3-3}

\begin{figure}[ht!]
\centering
\includegraphics*[clip,width=60mm]{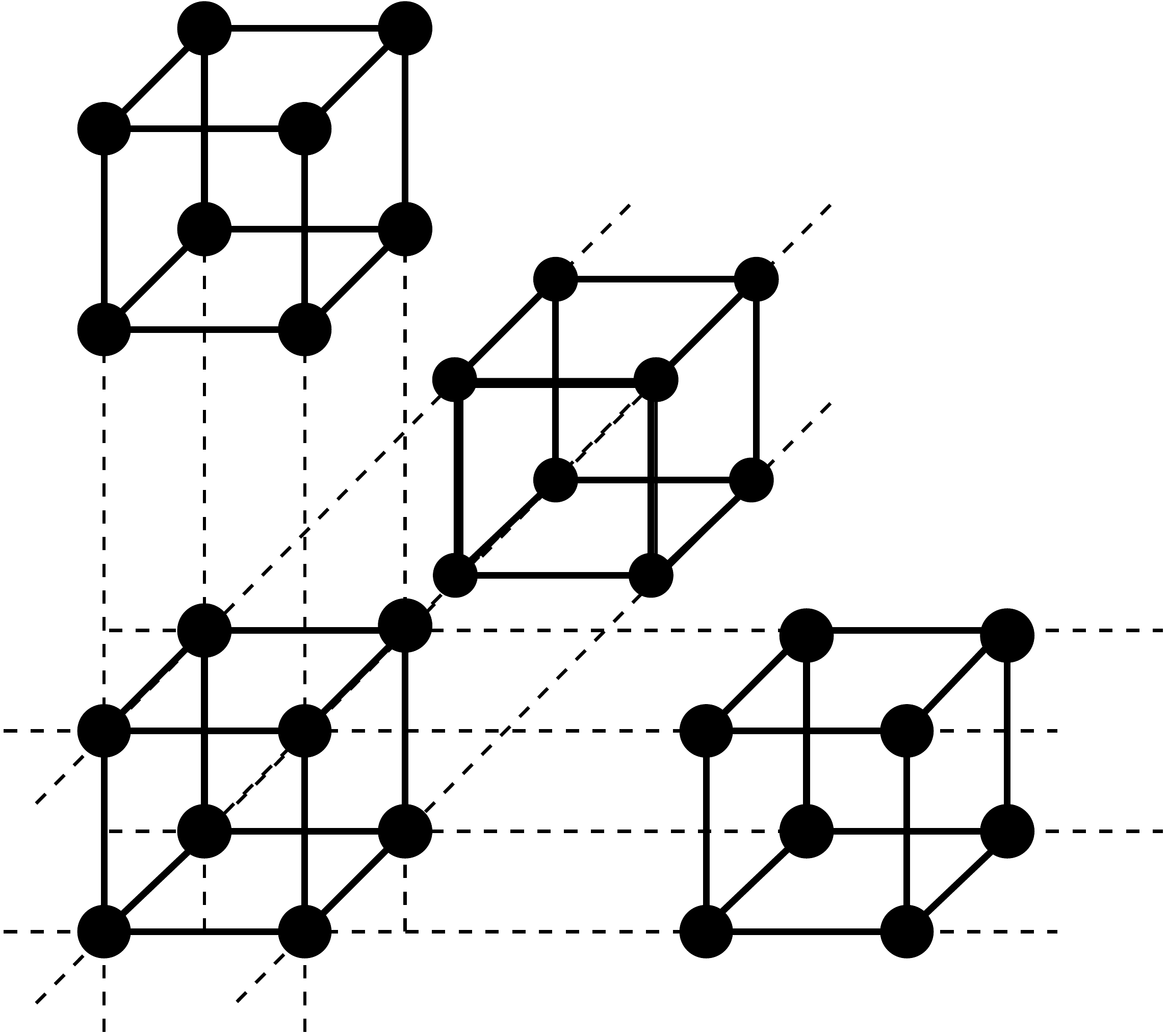}
\caption{Schematic structure of the
  investigated three-dimensional bipartite lattice; solid bonds
  depict interactions $J_1$, dashed bonds $J_2$.}   
\label{intermolecular-f-6}
\end{figure}

For the three-dimensional case we choose a simple cubic lattice
where the molecular unit is given by cubes as sketched in
\figref{intermolecular-f-6}. An isolated cube of spins $s=1/2$
and af bonds along the edges shows four magnetization steps at
low temperature which result from the successive level crossings
of the lowest states with total spin $S=0, 1, 2, 3, 4$. The
simple cubic lattice with $J_2/J_1=1$ on the other hand is a
system with long range order at $T>0$, a property which the lower
dimensional systems did not show in accordance with the theorem
of Mermin and Wagner.\cite{MeW:PRL66}

\begin{figure}[ht!]
\centering
\includegraphics*[clip,width=60mm]{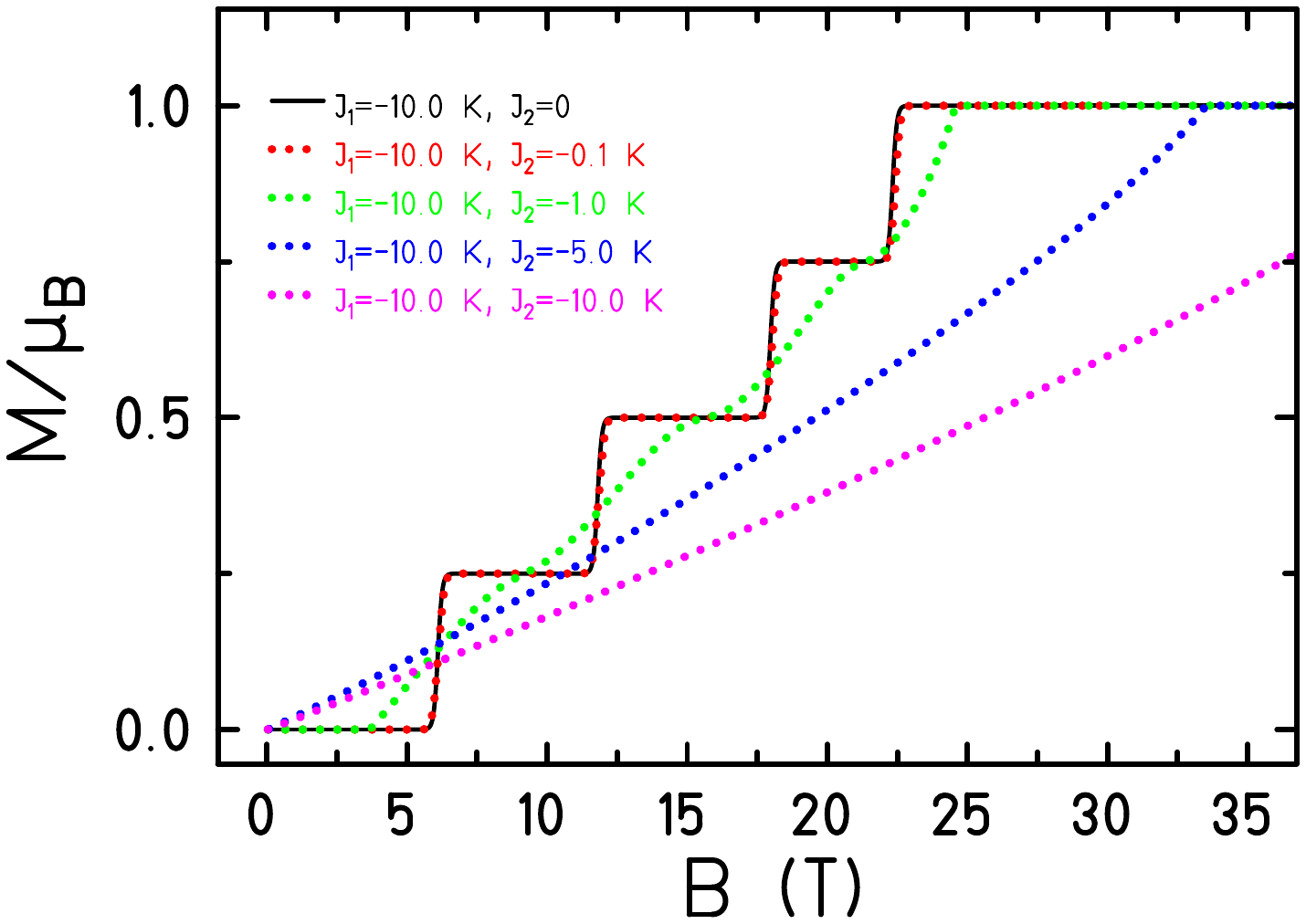}
\caption{(Color online) Low-temperature magnetization of the
  three-dimensional spin system shown in
  \figref{intermolecular-f-6} for various interdimer
  couplings $J_2$ and $T=0.1$~K.}   
\label{intermolecular-f-7}
\end{figure}

Looking at the magnetization in \figref{intermolecular-f-7} one
immediately realizes that already a rather small intermolecular
interaction of 10~\% suffices to wash out the magnetization steps
of the spin cube. It is important to keep in mind that the cube
has almost the same singlet-triplet gap as dimer and square, so
the effect is not thermal. We thus speculate that the
dimensionality of the embedding structure, here three, is
responsible for the quick disappearance of the molecular
fingerprints with increasing intermolecular interaction.

\begin{figure}[ht!]
\centering
\includegraphics*[clip,width=60mm]{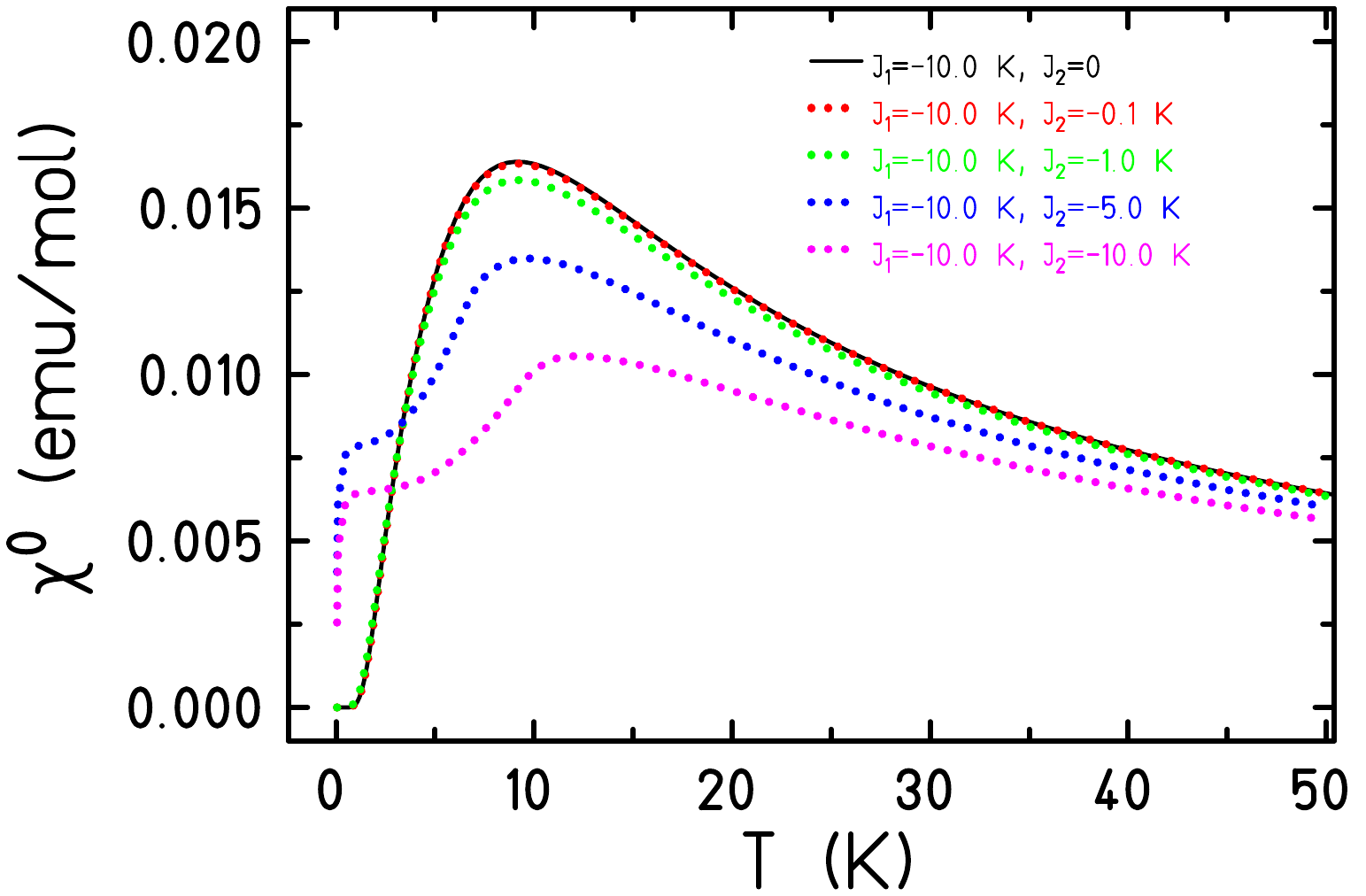}

\includegraphics*[clip,width=60mm]{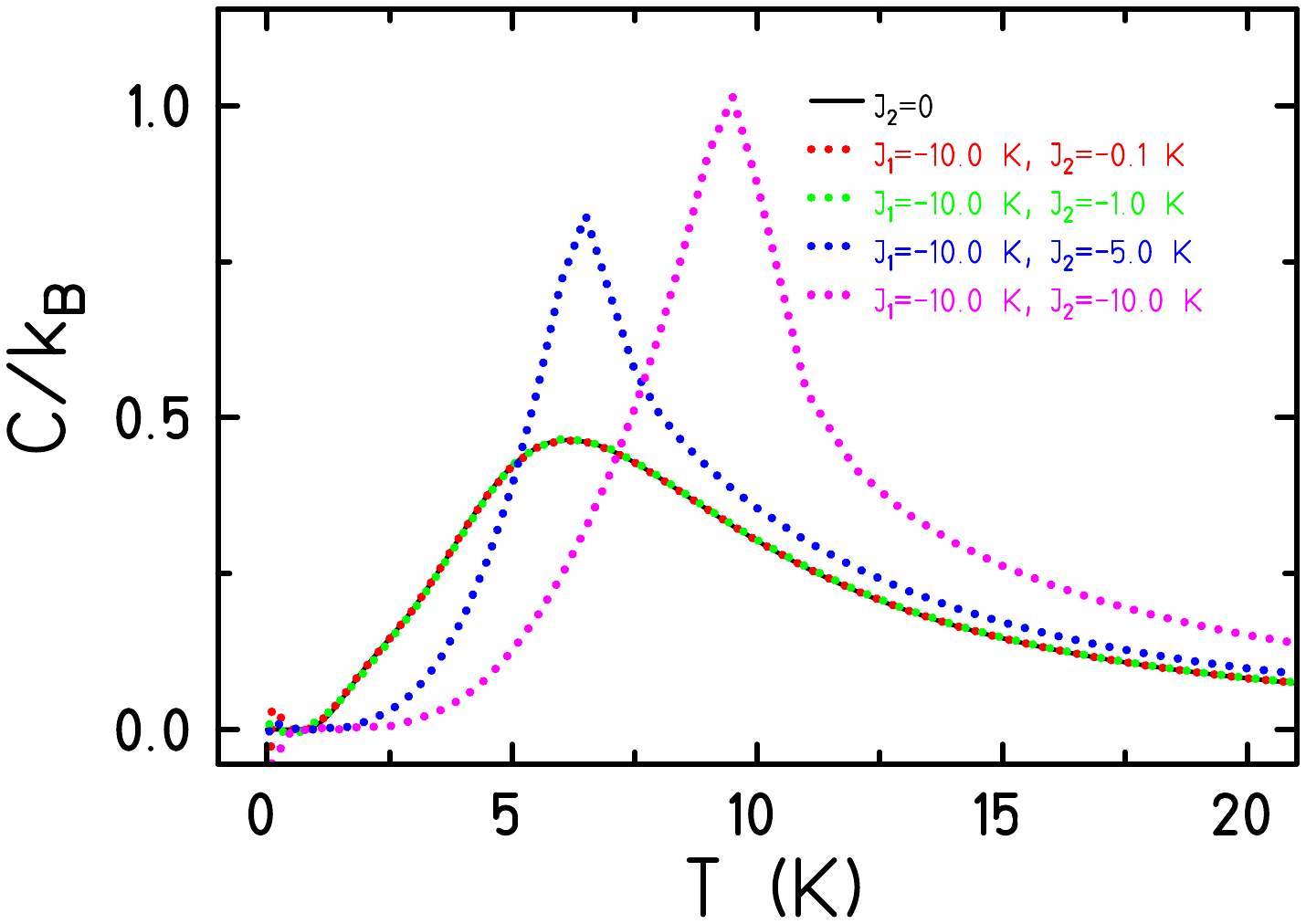}
\caption{(Color online) Zero-field susceptibility and specific heat of the
  three-dimensional spin system shown in
  \figref{intermolecular-f-6} for various interdimer
  couplings $J_2$ and $B=0$.}   
\label{intermolecular-f-8}
\end{figure}

Although the magnetization is already drastically altered by
10~\% intermolecular interactions, the temperature dependence
of the susceptibility does not show much deviation in this case,
compare \figref{intermolecular-f-8}. The same holds for the
specific heat. These functions are modified only for larger
intermolecular interactions in accord with the one- and
two-dimensional cases. The peaks of the specific heat for
$J_2/J_1=0.5$ and $J_2/J_1=1.0$ mark phase transitions to
three-dimensional ordered phases -- they correspond exactly to
those shown in Ref.~\onlinecite{SSS:PRB03}.

\section{Dimers in various dimensions}
\label{sec-4}

In a second setup we kept the molecular unit fixed as a
dimer and varied the dimension of the embedding. The
one-dimensional case remains the same. The two-dimensional case
can be derived from \figref{intermolecular-f-1}~(b) by replacing
all (thick) vertical $J_1$-bonds by (dashed) $J_2$-bonds. For
the three-dimensional case the two-dimensional lattices are stacked on top of
each other with $J_2$-bonds in between. Thus each spin is
connected by one $J_1$-bond and one, three, and five $J_2$-bonds
for the one-, two-, and three-dimensional case, respectively.

\begin{figure}[ht!]
\centering
\includegraphics*[clip,width=60mm]{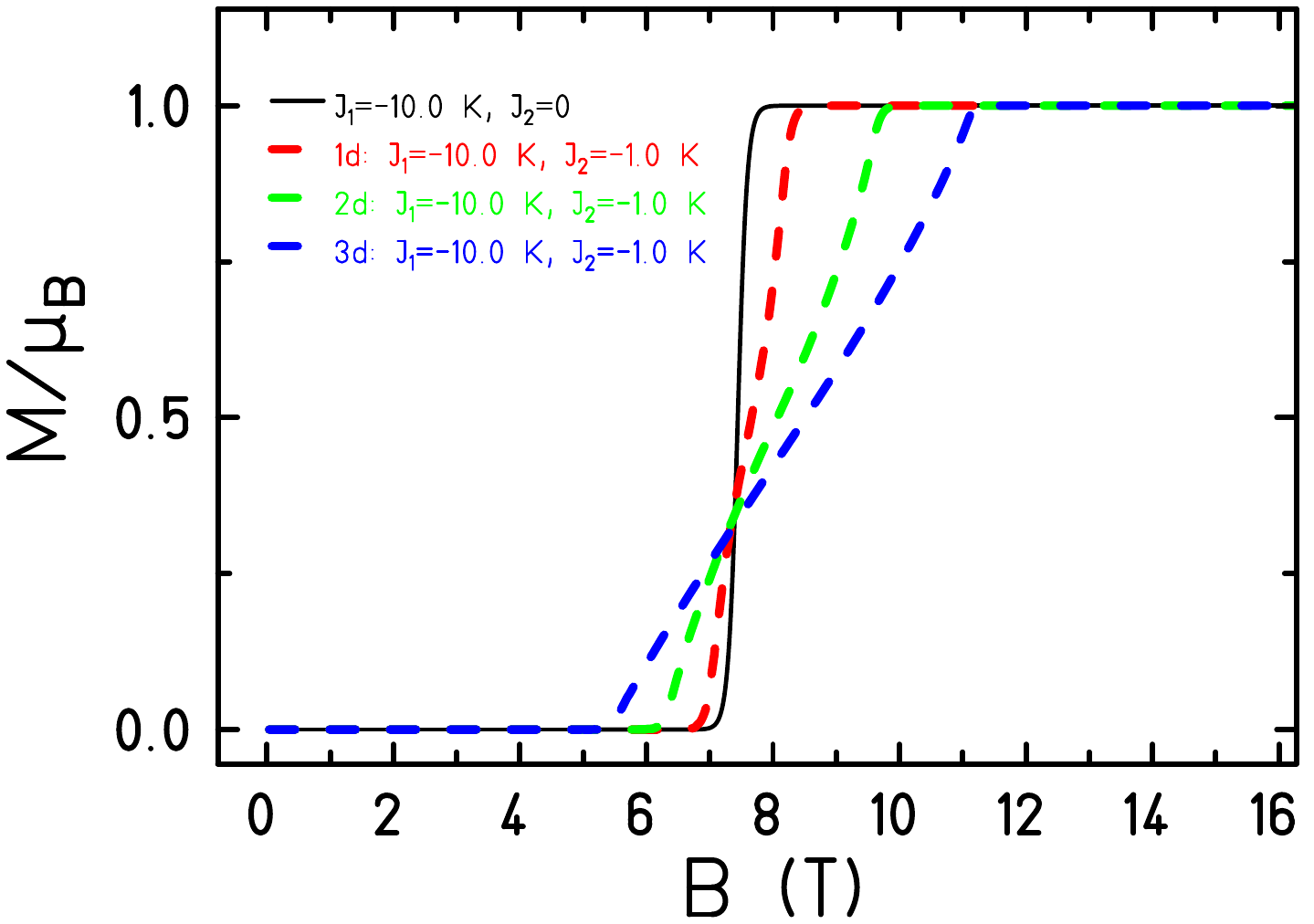}
\caption{(Color online) Low-temperature magnetization of dimers
  in one-, two and three-dimensional arrangements for
  $J_2=1.0$~K and $T=0.1$~K.}    
\label{intermolecular-f-9}
\end{figure}

\begin{figure}[ht!]
\centering
\includegraphics*[clip,width=60mm]{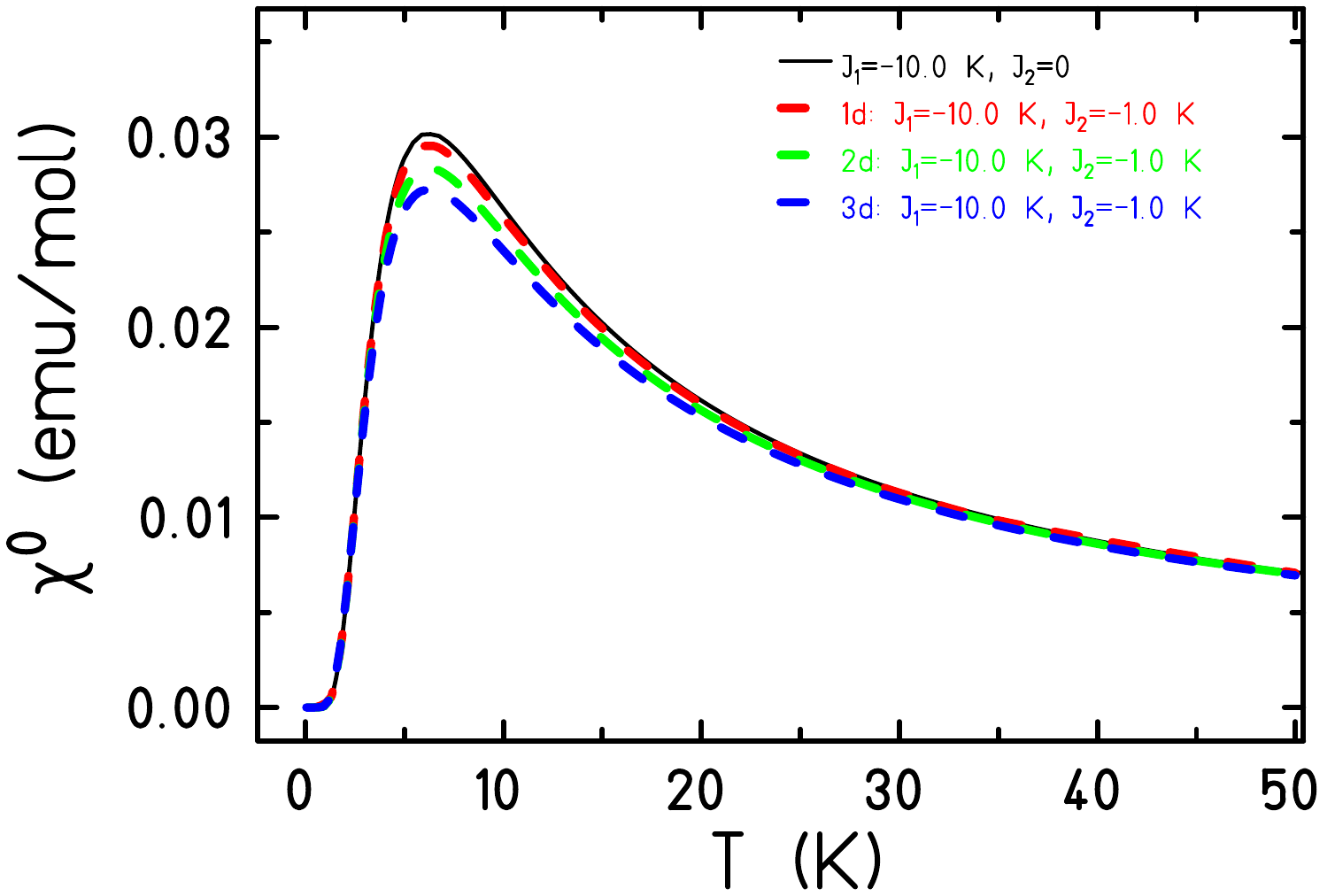}

\includegraphics*[clip,width=60mm]{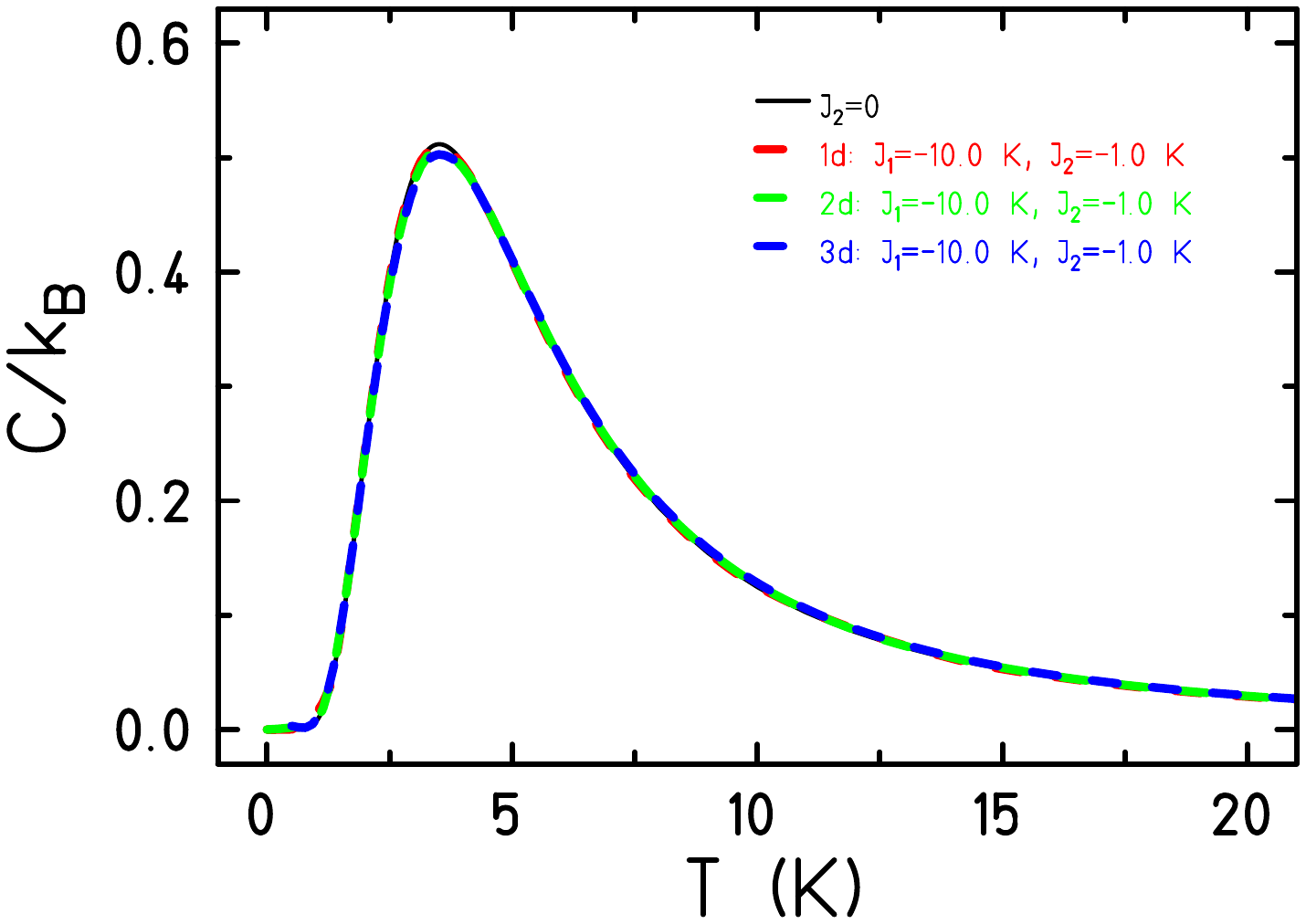}
\caption{(Color online) Zero-field susceptibility and specific
  heat of dimers   in one-, two and three-dimensional
  arrangements for   $J_2=1.0$~K and $T=0.1$~K.}   
\label{intermolecular-f-10}
\end{figure}

For the following investigation $J_2/J_1=0.1$ as well as the
temperature were kept constant. As can be clearly seen in
\figref{intermolecular-f-9} the magnetization step is more
strongly washed out with increasing dimensionality. The
influence on the temperature dependence of both susceptibility
as well as specific heat is again weak, see
\figref{intermolecular-f-10}. 

\section{Comparison with J-strain}
\label{sec-5}

Finally, as a supplement to the presented investigations, we
would like to discuss the question whether a similar
modification of observables could stem from J-strain. The
assumption of strain, for instance g-strain, is not unusual for
instance when modeling EPR lines. J-strain, i.e. a distribution of
$J$ values about a mean was used in several theoretical models,
see e.g. Refs.~\onlinecite{SPK:PRB08,SFF:JPCM:10,PPS:CPC15}. The
effect of J-strain is rather similar to that of intermolecular
interactions: magnetization steps are smeared out, and
susceptibility as well as specific heat as functions of
temperature are not much altered.

\begin{figure}[ht!]
\centering
\includegraphics*[clip,width=60mm]{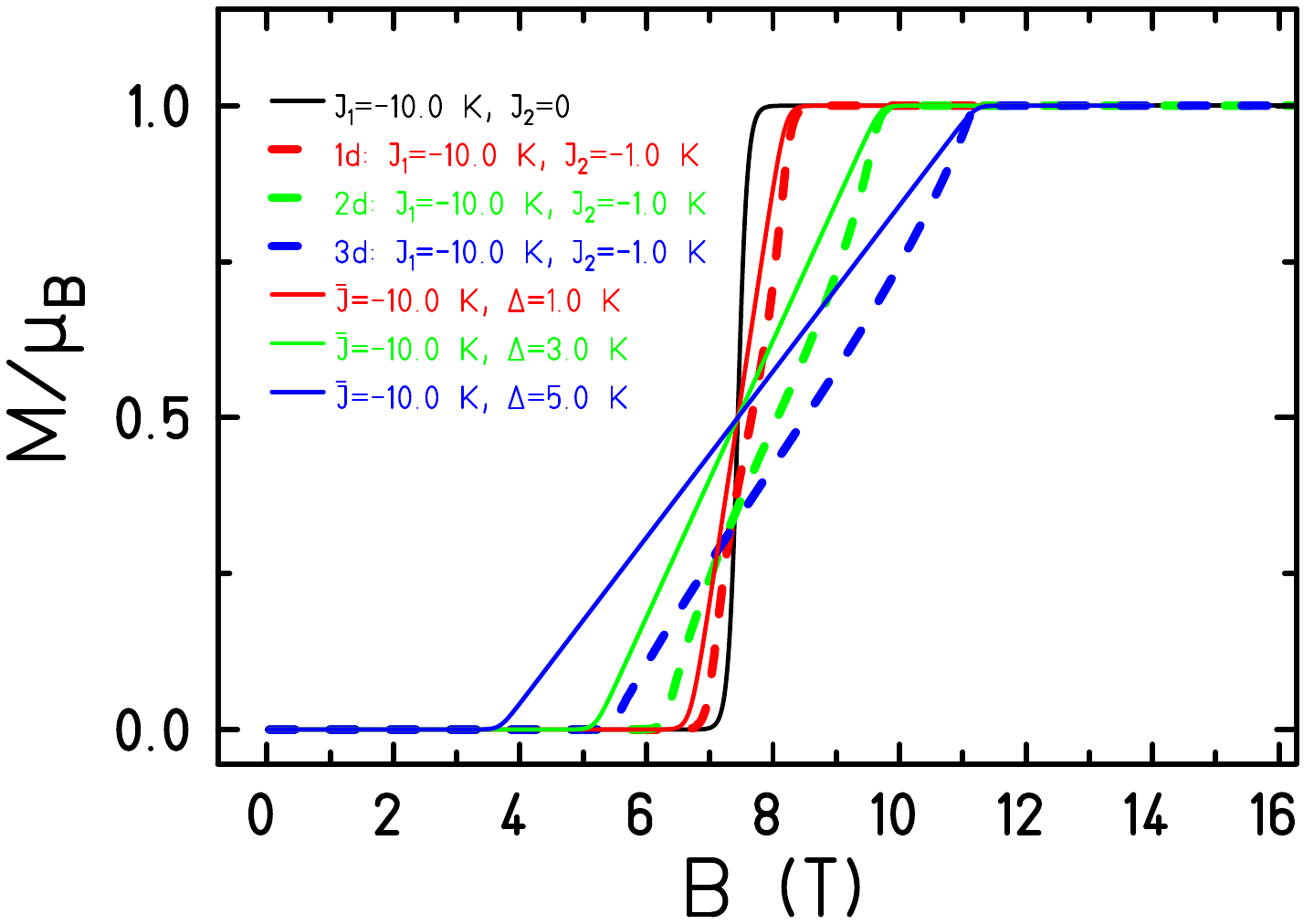}
\caption{(Color online) Low-temperature magnetization of dimers
  in one-, two and three-dimensional arrangements for
  $J_2=1.0$~K and $T=0.1$~K (dashes) compared to isolated dimers
  with a J-strain of $\Delta = 1.0, 3.0, 5.0$~K (solid curves),
  respectively.}
\label{intermolecular-f-11}
\end{figure}

In the following we present an investigation in which
independent dimers with a flat distribution of $J_1$-values in the
interval $[\bar{J}-\Delta,\bar{J}+\Delta ]$ have been
simulated. $\Delta$ was chosen such, that the saturation field
for the three cases discussed in section \ref{sec-4} is met.
Figure~\xref{intermolecular-f-11} shows a comparison of the
magnetization of a single dimer (black solid curve), of dimers
with intermolecular interactions in one, two, and three space
dimensions (dashed curves) as well as of dimers with J-strain
according to the flat distribution (solid colored curves). One
immediately realizes that the functional form of the
magnetization curve with J-strain is different from the behavior
under the influence of intermolecular interactions. Although the
saturation field is met by tuning $\Delta$ appropriately, the
onset of the magnetization curves happens already at smaller
fields. In addition, at the field value where the magnetization
step happens for the unperturbed dimer, the magnetization curves
of dimers with J-strain cross at half the step height whereas for
intermolecular interactions the magnetization curves cross at a
lower magnetization. Overall, the magnetization curves for
J-strain are symmetric about the crossing field value. This
would also hold if another (more realistic, but also symmetric
about $\bar{J}$) Gaussian
distribution of $J_1$ values would have been taken. Intermolecular
interactions on the contrary seem to lead to magnetization
curves, that do not show any symmetry with respect to the
original crossing field.

\begin{figure}[ht!]
\centering
\includegraphics*[clip,width=60mm]{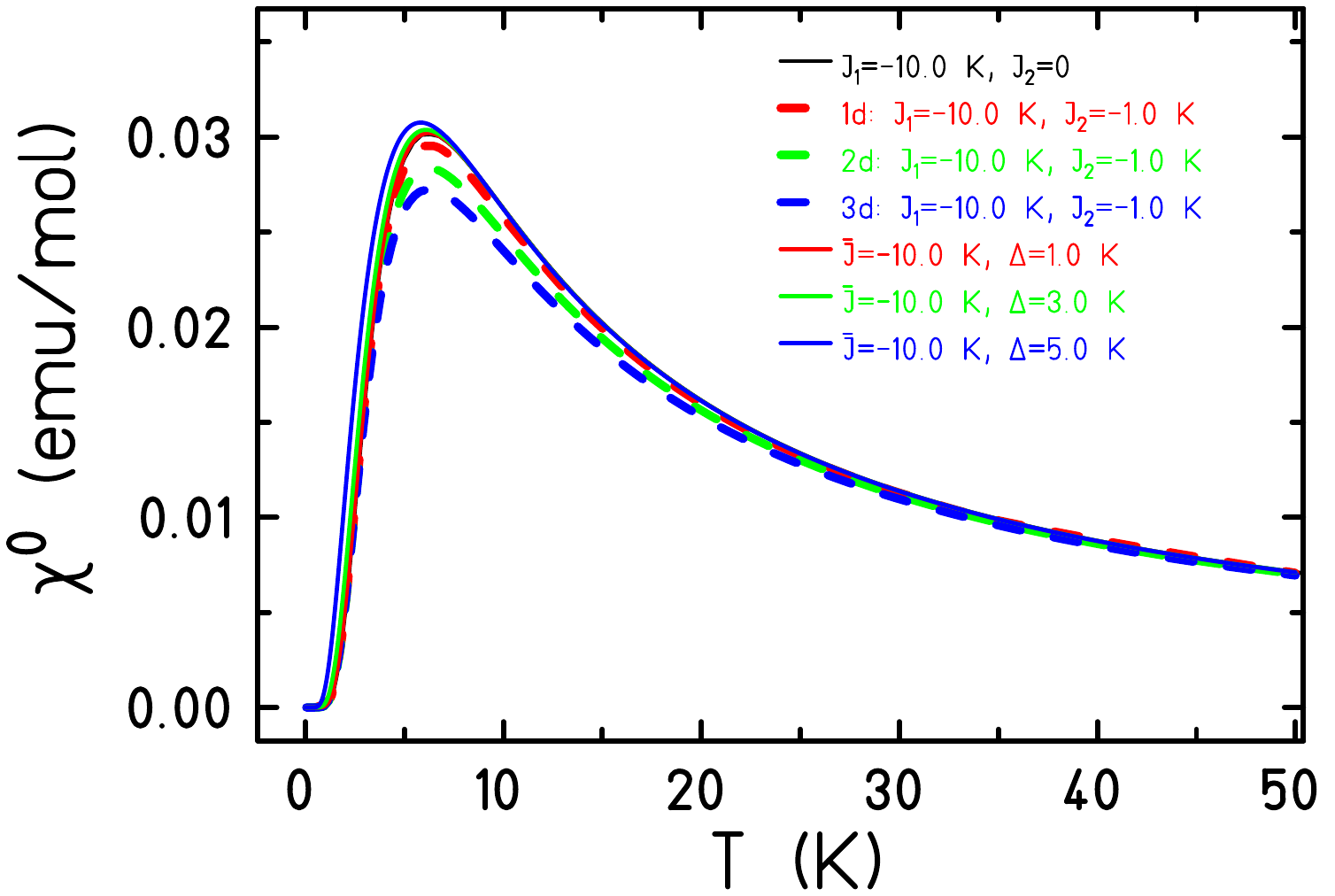}

\includegraphics*[clip,width=60mm]{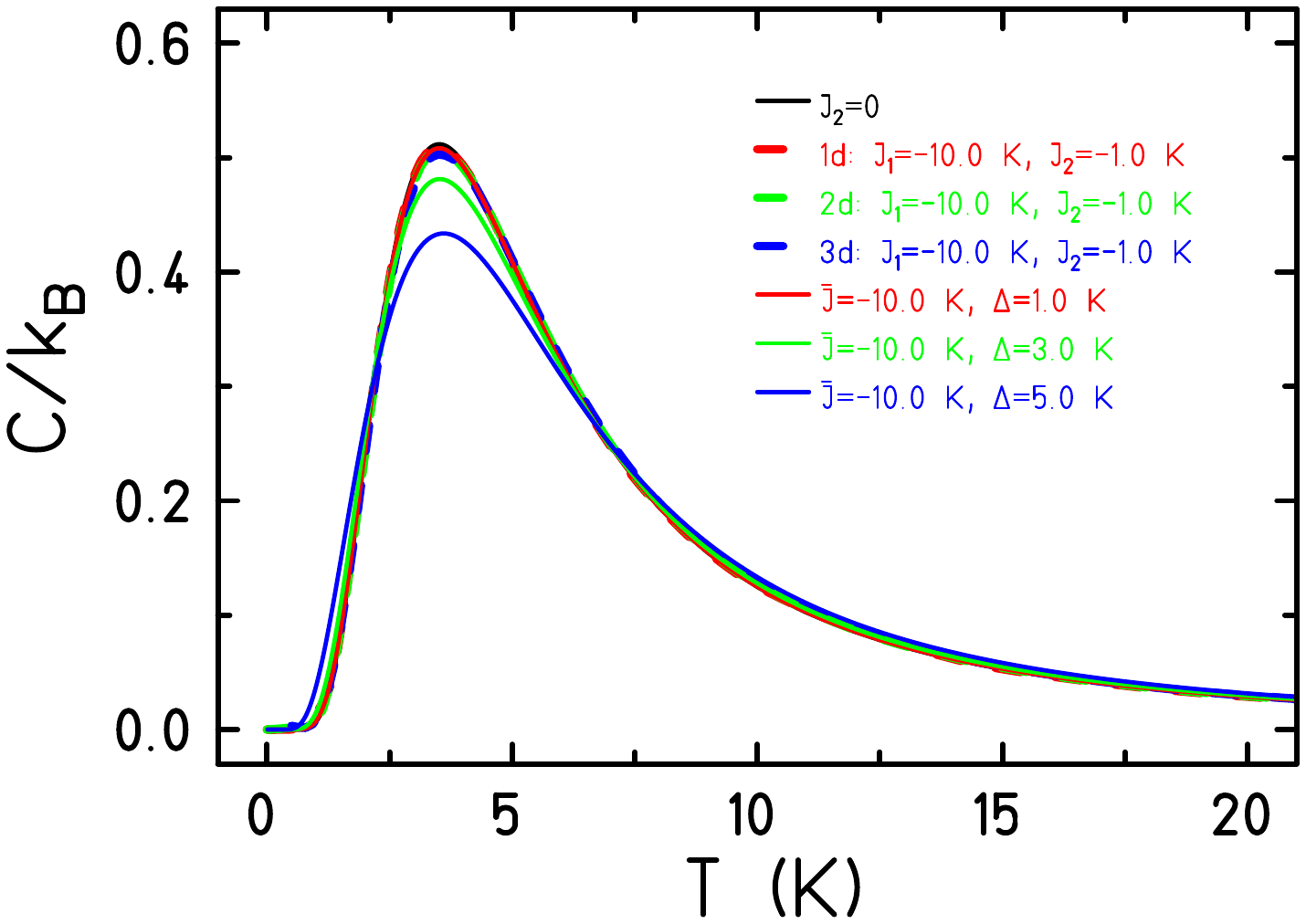}
\caption{(Color online) Zero-field susceptibility and specific
  heat of dimers in one-, two and three-dimensional
  arrangements for $J_2=1.0$~K and $T=0.1$~K (dashes) compared
  to isolated dimers with a J-strain of $\Delta = 1.0, 3.0,
  5.0$~K (solid curves), respectively.}
\label{intermolecular-f-12}
\end{figure}

Figure~\xref{intermolecular-f-12} demonstrates that somewhat
contrary to the findings of section \ref{sec-4} now the
susceptibility is only very weakly altered whereas the specific
heat is more drastically modified especially for the case of the 
largest J-strain.

\section{Summary and Outlook}
\label{sec-6}

We investigated the question how intermolecular interactions
influence magnetic observables for small (molecular) magnetic
units. In particular we investigated for certain bipartite
configurations how large the intermolecular interaction needs to
be compared to the intramolecular interaction in order to mask
the molecular behavior. It could be demonstrated that the
various static magnetic observables reflect intermolecular
interactions differently: the low-temperature magnetization
turned out to be most sensitive, since the appearance of
magnetization steps appears to be fragile. In addition
dimensionality plays a role. With increasing space
dimensionality of the intermolecular coupling the effect of
masking molecular properties happens for smaller intermolecular
coupling. Finally we discussed briefly whether similar
modifications of observables could be misinterpreted as
J-strain. We pointed out, that certain features of the
observables are different in the two scenarios, so that with
good quality of experimental data a discrimination should be
possible.

\section*{Acknowledgment}

This work was supported by the Deutsche Forschungsgemeinschaft (DFG
SCHN 615/20-1). I would like to thank Arzhang Ardavan, Stephen
Blundell, Marco Evangelisti, Andreas Honecker, Franziska
Kirschner, Hiroyuki Nojiri and Johannes Richter for valuable
discussions.

%


\begin{thebibliography}{61}%
\makeatletter
\providecommand \@ifxundefined [1]{%
 \@ifx{#1\undefined}
}%
\providecommand \@ifnum [1]{%
 \ifnum #1\expandafter \@firstoftwo
 \else \expandafter \@secondoftwo
 \fi
}%
\providecommand \@ifx [1]{%
 \ifx #1\expandafter \@firstoftwo
 \else \expandafter \@secondoftwo
 \fi
}%
\providecommand \natexlab [1]{#1}%
\providecommand \enquote  [1]{``#1''}%
\providecommand \bibnamefont  [1]{#1}%
\providecommand \bibfnamefont [1]{#1}%
\providecommand \citenamefont [1]{#1}%
\providecommand \href@noop [0]{\@secondoftwo}%
\providecommand \href [0]{\begingroup \@sanitize@url \@href}%
\providecommand \@href[1]{\@@startlink{#1}\@@href}%
\providecommand \@@href[1]{\endgroup#1\@@endlink}%
\providecommand \@sanitize@url [0]{\catcode `\\12\catcode `\$12\catcode
  `\&12\catcode `\#12\catcode `\^12\catcode `\_12\catcode `\%12\relax}%
\providecommand \@@startlink[1]{}%
\providecommand \@@endlink[0]{}%
\providecommand \url  [0]{\begingroup\@sanitize@url \@url }%
\providecommand \@url [1]{\endgroup\@href {#1}{\urlprefix }}%
\providecommand \urlprefix  [0]{URL }%
\providecommand \Eprint [0]{\href }%
\providecommand \doibase [0]{http://dx.doi.org/}%
\providecommand \selectlanguage [0]{\@gobble}%
\providecommand \bibinfo  [0]{\@secondoftwo}%
\providecommand \bibfield  [0]{\@secondoftwo}%
\providecommand \translation [1]{[#1]}%
\providecommand \BibitemOpen [0]{}%
\providecommand \bibitemStop [0]{}%
\providecommand \bibitemNoStop [0]{.\EOS\space}%
\providecommand \EOS [0]{\spacefactor3000\relax}%
\providecommand \BibitemShut  [1]{\csname bibitem#1\endcsname}%
\let\auto@bib@innerbib\@empty
\bibitem [{\citenamefont {Ardavan}\ \emph {et~al.}(2007)\citenamefont
  {Ardavan}, \citenamefont {Rival}, \citenamefont {Morton}, \citenamefont
  {Blundell}, \citenamefont {Tyryshkin}, \citenamefont {Timco},\ and\
  \citenamefont {Winpenny}}]{ARM:PRL07}%
  \BibitemOpen
  \bibfield  {author} {\bibinfo {author} {\bibfnamefont {A.}~\bibnamefont
  {Ardavan}}, \bibinfo {author} {\bibfnamefont {O.}~\bibnamefont {Rival}},
  \bibinfo {author} {\bibfnamefont {J.~J.~L.}\ \bibnamefont {Morton}}, \bibinfo
  {author} {\bibfnamefont {S.~J.}\ \bibnamefont {Blundell}}, \bibinfo {author}
  {\bibfnamefont {A.~M.}\ \bibnamefont {Tyryshkin}}, \bibinfo {author}
  {\bibfnamefont {G.~A.}\ \bibnamefont {Timco}}, \ and\ \bibinfo {author}
  {\bibfnamefont {R.~E.~P.}\ \bibnamefont {Winpenny}},\ }\enquote {\bibinfo
  {title} {Will Spin-Relaxation Times in Molecular Magnets Permit Quantum
  Information Processing?}}\ \href
  {http://link.aps.org/abstract/PRL/v98/e057201} {\bibfield  {journal}
  {\bibinfo  {journal} {Phys. Rev. Lett.}\ }\textbf {\bibinfo {volume} {98}},\
  \bibinfo {pages} {057201} (\bibinfo {year} {2007})}\BibitemShut {NoStop}%
\bibitem [{\citenamefont {Candini}\ \emph {et~al.}(2010)\citenamefont
  {Candini}, \citenamefont {Lorusso}, \citenamefont {Troiani}, \citenamefont
  {Ghirri}, \citenamefont {Carretta}, \citenamefont {Santini}, \citenamefont
  {Amoretti}, \citenamefont {Muryn}, \citenamefont {Tuna}, \citenamefont
  {Timco}, \citenamefont {McInnes}, \citenamefont {Winpenny}, \citenamefont
  {Wernsdorfer},\ and\ \citenamefont {Affronte}}]{CLT:PRL10}%
  \BibitemOpen
  \bibfield  {author} {\bibinfo {author} {\bibfnamefont {A.}~\bibnamefont
  {Candini}}, \bibinfo {author} {\bibfnamefont {G.}~\bibnamefont {Lorusso}},
  \bibinfo {author} {\bibfnamefont {F.}~\bibnamefont {Troiani}}, \bibinfo
  {author} {\bibfnamefont {A.}~\bibnamefont {Ghirri}}, \bibinfo {author}
  {\bibfnamefont {S.}~\bibnamefont {Carretta}}, \bibinfo {author}
  {\bibfnamefont {P.}~\bibnamefont {Santini}}, \bibinfo {author} {\bibfnamefont
  {G.}~\bibnamefont {Amoretti}}, \bibinfo {author} {\bibfnamefont
  {C.}~\bibnamefont {Muryn}}, \bibinfo {author} {\bibfnamefont
  {F.}~\bibnamefont {Tuna}}, \bibinfo {author} {\bibfnamefont {G.}~\bibnamefont
  {Timco}}, \bibinfo {author} {\bibfnamefont {E.~J.~L.}\ \bibnamefont
  {McInnes}}, \bibinfo {author} {\bibfnamefont {R.~E.~P.}\ \bibnamefont
  {Winpenny}}, \bibinfo {author} {\bibfnamefont {W.}~\bibnamefont
  {Wernsdorfer}}, \ and\ \bibinfo {author} {\bibfnamefont {M.}~\bibnamefont
  {Affronte}},\ }\enquote {\bibinfo {title} {Entanglement in Supramolecular
  Spin Systems of Two Weakly Coupled Antiferromagnetic Rings
  (Purple-Cr$_{7}$Ni)},}\ \href {\doibase 10.1103/PhysRevLett.104.037203}
  {\bibfield  {journal} {\bibinfo  {journal} {Phys. Rev. Lett.}\ }\textbf
  {\bibinfo {volume} {104}},\ \bibinfo {pages} {037203} (\bibinfo {year}
  {2010})}\BibitemShut {NoStop}%
\bibitem [{\citenamefont {Chiesa}\ \emph {et~al.}(2014)\citenamefont {Chiesa},
  \citenamefont {Gerace}, \citenamefont {Troiani}, \citenamefont {Amoretti},
  \citenamefont {Santini},\ and\ \citenamefont {Carretta}}]{CGT:PRA14}%
  \BibitemOpen
  \bibfield  {author} {\bibinfo {author} {\bibfnamefont {A.}~\bibnamefont
  {Chiesa}}, \bibinfo {author} {\bibfnamefont {D.}~\bibnamefont {Gerace}},
  \bibinfo {author} {\bibfnamefont {F.}~\bibnamefont {Troiani}}, \bibinfo
  {author} {\bibfnamefont {G.}~\bibnamefont {Amoretti}}, \bibinfo {author}
  {\bibfnamefont {P.}~\bibnamefont {Santini}}, \ and\ \bibinfo {author}
  {\bibfnamefont {S.}~\bibnamefont {Carretta}},\ }\enquote {\bibinfo {title}
  {Robustness of quantum gates with hybrid spin-photon qubits in
  superconducting resonators},}\ \href {\doibase 10.1103/PhysRevA.89.052308}
  {\bibfield  {journal} {\bibinfo  {journal} {Phys. Rev. A}\ }\textbf {\bibinfo
  {volume} {89}},\ \bibinfo {pages} {052308} (\bibinfo {year}
  {2014})}\BibitemShut {NoStop}%
\bibitem [{\citenamefont {Pineda}\ \emph {et~al.}(2014)\citenamefont {Pineda},
  \citenamefont {Chilton}, \citenamefont {Marx}, \citenamefont {D{\"o}rfel},
  \citenamefont {Sells}, \citenamefont {Neugebauer}, \citenamefont {Jiang},
  \citenamefont {Collison}, \citenamefont {van Slageren}, \citenamefont
  {McInnes},\ and\ \citenamefont {Winpenny}}]{PCM:NC14}%
  \BibitemOpen
  \bibfield  {author} {\bibinfo {author} {\bibfnamefont {E.~M.}\ \bibnamefont
  {Pineda}}, \bibinfo {author} {\bibfnamefont {N.~F.}\ \bibnamefont {Chilton}},
  \bibinfo {author} {\bibfnamefont {R.}~\bibnamefont {Marx}}, \bibinfo {author}
  {\bibfnamefont {M.}~\bibnamefont {D{\"o}rfel}}, \bibinfo {author}
  {\bibfnamefont {D.~O.}\ \bibnamefont {Sells}}, \bibinfo {author}
  {\bibfnamefont {P.}~\bibnamefont {Neugebauer}}, \bibinfo {author}
  {\bibfnamefont {S.-D.}\ \bibnamefont {Jiang}}, \bibinfo {author}
  {\bibfnamefont {D.}~\bibnamefont {Collison}}, \bibinfo {author}
  {\bibfnamefont {J.}~\bibnamefont {van Slageren}}, \bibinfo {author}
  {\bibfnamefont {E.~J.}\ \bibnamefont {McInnes}}, \ and\ \bibinfo {author}
  {\bibfnamefont {R.~E.}\ \bibnamefont {Winpenny}},\ }\enquote {\bibinfo
  {title} {Direct measurement of dysprosium(III)$\cdots$dysprosium(III)
  interactions in a single-molecule magnet},}\ \href {\doibase
  10.1038/ncomms6243} {\bibfield  {journal} {\bibinfo  {journal} {Nat.
  Commun.}\ }\textbf {\bibinfo {volume} {5}},\ \bibinfo {pages} {5243}
  (\bibinfo {year} {2014})}\BibitemShut {NoStop}%
\bibitem [{\citenamefont {Lancaster}\ \emph {et~al.}(2014)\citenamefont
  {Lancaster}, \citenamefont {Goddard}, \citenamefont {Blundell}, \citenamefont
  {Foronda}, \citenamefont {Ghannadzadeh}, \citenamefont {M\"oller},
  \citenamefont {Baker}, \citenamefont {Pratt}, \citenamefont {Baines},
  \citenamefont {Huang}, \citenamefont {Wosnitza}, \citenamefont {McDonald},
  \citenamefont {Modic}, \citenamefont {Singleton}, \citenamefont {Topping},
  \citenamefont {Beale}, \citenamefont {Xiao}, \citenamefont {Schlueter},
  \citenamefont {Barton}, \citenamefont {Cabrera}, \citenamefont {Carreiro},
  \citenamefont {Tran},\ and\ \citenamefont {Manson}}]{LGB:PRL14}%
  \BibitemOpen
  \bibfield  {author} {\bibinfo {author} {\bibfnamefont {T.}~\bibnamefont
  {Lancaster}}, \bibinfo {author} {\bibfnamefont {P.~A.}\ \bibnamefont
  {Goddard}}, \bibinfo {author} {\bibfnamefont {S.~J.}\ \bibnamefont
  {Blundell}}, \bibinfo {author} {\bibfnamefont {F.~R.}\ \bibnamefont
  {Foronda}}, \bibinfo {author} {\bibfnamefont {S.}~\bibnamefont
  {Ghannadzadeh}}, \bibinfo {author} {\bibfnamefont {J.~S.}\ \bibnamefont
  {M\"oller}}, \bibinfo {author} {\bibfnamefont {P.~J.}\ \bibnamefont {Baker}},
  \bibinfo {author} {\bibfnamefont {F.~L.}\ \bibnamefont {Pratt}}, \bibinfo
  {author} {\bibfnamefont {C.}~\bibnamefont {Baines}}, \bibinfo {author}
  {\bibfnamefont {L.}~\bibnamefont {Huang}}, \bibinfo {author} {\bibfnamefont
  {J.}~\bibnamefont {Wosnitza}}, \bibinfo {author} {\bibfnamefont {R.~D.}\
  \bibnamefont {McDonald}}, \bibinfo {author} {\bibfnamefont {K.~A.}\
  \bibnamefont {Modic}}, \bibinfo {author} {\bibfnamefont {J.}~\bibnamefont
  {Singleton}}, \bibinfo {author} {\bibfnamefont {C.~V.}\ \bibnamefont
  {Topping}}, \bibinfo {author} {\bibfnamefont {T.~A.~W.}\ \bibnamefont
  {Beale}}, \bibinfo {author} {\bibfnamefont {F.}~\bibnamefont {Xiao}},
  \bibinfo {author} {\bibfnamefont {J.~A.}\ \bibnamefont {Schlueter}}, \bibinfo
  {author} {\bibfnamefont {A.~M.}\ \bibnamefont {Barton}}, \bibinfo {author}
  {\bibfnamefont {R.~D.}\ \bibnamefont {Cabrera}}, \bibinfo {author}
  {\bibfnamefont {K.~E.}\ \bibnamefont {Carreiro}}, \bibinfo {author}
  {\bibfnamefont {H.~E.}\ \bibnamefont {Tran}}, \ and\ \bibinfo {author}
  {\bibfnamefont {J.~L.}\ \bibnamefont {Manson}},\ }\enquote {\bibinfo {title}
  {Controlling Magnetic Order and Quantum Disorder in Molecule-Based
  Magnets},}\ \href {\doibase 10.1103/PhysRevLett.112.207201} {\bibfield
  {journal} {\bibinfo  {journal} {Phys. Rev. Lett.}\ }\textbf {\bibinfo
  {volume} {112}},\ \bibinfo {pages} {207201} (\bibinfo {year}
  {2014})}\BibitemShut {NoStop}%
\bibitem [{\citenamefont {Lapasar}\ \emph {et~al.}(2014)\citenamefont
  {Lapasar}, \citenamefont {Kasamatsu}, \citenamefont {Chormaic}, \citenamefont
  {Takui}, \citenamefont {Kondo}, \citenamefont {Nakahara},\ and\ \citenamefont
  {Ohmi}}]{LKC:JPSJ14}%
  \BibitemOpen
  \bibfield  {author} {\bibinfo {author} {\bibfnamefont {E.~H.}\ \bibnamefont
  {Lapasar}}, \bibinfo {author} {\bibfnamefont {K.}~\bibnamefont {Kasamatsu}},
  \bibinfo {author} {\bibfnamefont {S.~N.}\ \bibnamefont {Chormaic}}, \bibinfo
  {author} {\bibfnamefont {T.}~\bibnamefont {Takui}}, \bibinfo {author}
  {\bibfnamefont {Y.}~\bibnamefont {Kondo}}, \bibinfo {author} {\bibfnamefont
  {M.}~\bibnamefont {Nakahara}}, \ and\ \bibinfo {author} {\bibfnamefont
  {T.}~\bibnamefont {Ohmi}},\ }\enquote {\bibinfo {title} {Two-Qubit Gate
  Operation on Selected Nearest-Neighbor Neutral Atom Qubits},}\ \href
  {\doibase 10.7566/JPSJ.83.044005} {\bibfield  {journal} {\bibinfo  {journal}
  {J. Phys. Soc. Jpn.}\ }\textbf {\bibinfo {volume} {83}},\ \bibinfo {pages}
  {044005} (\bibinfo {year} {2014})}\BibitemShut {NoStop}%
\bibitem [{\citenamefont {Yamamoto}\ \emph {et~al.}(2015)\citenamefont
  {Yamamoto}, \citenamefont {Nakazawa}, \citenamefont {Sugisaki}, \citenamefont
  {Sato}, \citenamefont {Toyota}, \citenamefont {Shiomi},\ and\ \citenamefont
  {Takui}}]{YNS:PCCP15}%
  \BibitemOpen
  \bibfield  {author} {\bibinfo {author} {\bibfnamefont {S.}~\bibnamefont
  {Yamamoto}}, \bibinfo {author} {\bibfnamefont {S.}~\bibnamefont {Nakazawa}},
  \bibinfo {author} {\bibfnamefont {K.}~\bibnamefont {Sugisaki}}, \bibinfo
  {author} {\bibfnamefont {K.}~\bibnamefont {Sato}}, \bibinfo {author}
  {\bibfnamefont {K.}~\bibnamefont {Toyota}}, \bibinfo {author} {\bibfnamefont
  {D.}~\bibnamefont {Shiomi}}, \ and\ \bibinfo {author} {\bibfnamefont
  {T.}~\bibnamefont {Takui}},\ }\enquote {\bibinfo {title} {Adiabatic quantum
  computing with spin qubits hosted by molecules},}\ \href {\doibase
  10.1039/C4CP04744C} {\bibfield  {journal} {\bibinfo  {journal} {Phys. Chem.
  Chem. Phys.}\ }\textbf {\bibinfo {volume} {17}},\ \bibinfo {pages} {2742}
  (\bibinfo {year} {2015})}\BibitemShut {NoStop}%
\bibitem [{\citenamefont {Oguchi}(1964)}]{Ogu:PR64}%
  \BibitemOpen
  \bibfield  {author} {\bibinfo {author} {\bibfnamefont {T.}~\bibnamefont
  {Oguchi}},\ }\enquote {\bibinfo {title} {Exchange Interactions in
  Cu${(\mathrm{N}{\mathrm{H}}_{3})}_{4}$S${\mathrm{O}}_{4}$\ifmmode\cdot\else\textperiodcentered\fi{}${\mathrm{H}}_{2}$O},}\
  \href {\doibase 10.1103/PhysRev.133.A1098} {\bibfield  {journal} {\bibinfo
  {journal} {Phys. Rev.}\ }\textbf {\bibinfo {volume} {133}},\ \bibinfo {pages}
  {A1098} (\bibinfo {year} {1964})}\BibitemShut {NoStop}%
\bibitem [{\citenamefont {Tachiki}\ and\ \citenamefont
  {Yamada}(1970)}]{TaY:JPSJ70}%
  \BibitemOpen
  \bibfield  {author} {\bibinfo {author} {\bibfnamefont {M.}~\bibnamefont
  {Tachiki}}\ and\ \bibinfo {author} {\bibfnamefont {T.}~\bibnamefont
  {Yamada}},\ }\enquote {\bibinfo {title} {Spin Ordering in a Spin-Pair
  System},}\ \href {\doibase 10.1143/JPSJ.28.1413} {\bibfield  {journal}
  {\bibinfo  {journal} {J. Phys. Soc. Jpn.}\ }\textbf {\bibinfo {volume}
  {28}},\ \bibinfo {pages} {1413} (\bibinfo {year} {1970})}\BibitemShut
  {NoStop}%
\bibitem [{\citenamefont {Tachiki}\ \emph {et~al.}(1970)\citenamefont
  {Tachiki}, \citenamefont {Yamada},\ and\ \citenamefont
  {Maekawa}}]{TYM:JPSJ70}%
  \BibitemOpen
  \bibfield  {author} {\bibinfo {author} {\bibfnamefont {M.}~\bibnamefont
  {Tachiki}}, \bibinfo {author} {\bibfnamefont {T.}~\bibnamefont {Yamada}}, \
  and\ \bibinfo {author} {\bibfnamefont {S.}~\bibnamefont {Maekawa}},\
  }\enquote {\bibinfo {title} {Short Range Order of Spins in Cu(NO3)22.5H2O
  under Magnetic Field},}\ \href {\doibase 10.1143/JPSJ.29.663} {\bibfield
  {journal} {\bibinfo  {journal} {J. Phys. Soc. Jpn.}\ }\textbf {\bibinfo
  {volume} {29}},\ \bibinfo {pages} {663} (\bibinfo {year} {1970})}\BibitemShut
  {NoStop}%
\bibitem [{\citenamefont {Oguchi}\ and\ \citenamefont
  {Blume}(1981)}]{OgB:JPSJ81}%
  \BibitemOpen
  \bibfield  {author} {\bibinfo {author} {\bibfnamefont {T.}~\bibnamefont
  {Oguchi}}\ and\ \bibinfo {author} {\bibfnamefont {M.}~\bibnamefont {Blume}},\
  }\enquote {\bibinfo {title} {Theory of the Field Dependence of the Neel
  Temperature in Quasi One-Dimensional Antiferromagnets},}\ \href {\doibase
  10.1143/JPSJ.50.2547} {\bibfield  {journal} {\bibinfo  {journal} {J. Phys.
  Soc. Jpn.}\ }\textbf {\bibinfo {volume} {50}},\ \bibinfo {pages} {2547}
  (\bibinfo {year} {1981})}\BibitemShut {NoStop}%
\bibitem [{\citenamefont {Nakazawa}\ \emph {et~al.}(1992)\citenamefont
  {Nakazawa}, \citenamefont {Tamura}, \citenamefont {Shirakawa}, \citenamefont
  {Shiomi}, \citenamefont {Takahashi}, \citenamefont {Kinoshita},\ and\
  \citenamefont {Ishikawa}}]{NTS:PRB92}%
  \BibitemOpen
  \bibfield  {author} {\bibinfo {author} {\bibfnamefont {Y.}~\bibnamefont
  {Nakazawa}}, \bibinfo {author} {\bibfnamefont {M.}~\bibnamefont {Tamura}},
  \bibinfo {author} {\bibfnamefont {N.}~\bibnamefont {Shirakawa}}, \bibinfo
  {author} {\bibfnamefont {D.}~\bibnamefont {Shiomi}}, \bibinfo {author}
  {\bibfnamefont {M.}~\bibnamefont {Takahashi}}, \bibinfo {author}
  {\bibfnamefont {M.}~\bibnamefont {Kinoshita}}, \ and\ \bibinfo {author}
  {\bibfnamefont {M.}~\bibnamefont {Ishikawa}},\ }\enquote {\bibinfo {title}
  {Low-temperature magnetic properties of the ferromagnetic organic radical,
  \textit{p} -nitrophenyl nitronyl nitroxide},}\ \href {\doibase
  10.1103/PhysRevB.46.8906} {\bibfield  {journal} {\bibinfo  {journal} {Phys.
  Rev. B}\ }\textbf {\bibinfo {volume} {46}},\ \bibinfo {pages} {8906}
  (\bibinfo {year} {1992})}\BibitemShut {NoStop}%
\bibitem [{\citenamefont {Kodama}\ \emph {et~al.}(2002)\citenamefont {Kodama},
  \citenamefont {Takigawa}, \citenamefont {Horvatic}, \citenamefont {Berthier},
  \citenamefont {Kageyama}, \citenamefont {Ueda}, \citenamefont {Miyahara},
  \citenamefont {Becca},\ and\ \citenamefont {Mila}}]{KTH:Science02}%
  \BibitemOpen
  \bibfield  {author} {\bibinfo {author} {\bibfnamefont {K.}~\bibnamefont
  {Kodama}}, \bibinfo {author} {\bibfnamefont {M.}~\bibnamefont {Takigawa}},
  \bibinfo {author} {\bibfnamefont {M.}~\bibnamefont {Horvatic}}, \bibinfo
  {author} {\bibfnamefont {C.}~\bibnamefont {Berthier}}, \bibinfo {author}
  {\bibfnamefont {H.}~\bibnamefont {Kageyama}}, \bibinfo {author}
  {\bibfnamefont {Y.}~\bibnamefont {Ueda}}, \bibinfo {author} {\bibfnamefont
  {S.}~\bibnamefont {Miyahara}}, \bibinfo {author} {\bibfnamefont
  {F.}~\bibnamefont {Becca}}, \ and\ \bibinfo {author} {\bibfnamefont
  {F.}~\bibnamefont {Mila}},\ }\enquote {\bibinfo {title} {Magnetic
  Superstructure in the Two-Dimensional Quantum Antiferromagnet SrCu2(BO3)2},}\
  \href {\doibase 10.1126/science.1075045} {\bibfield  {journal} {\bibinfo
  {journal} {Science}\ }\textbf {\bibinfo {volume} {298}},\ \bibinfo {pages}
  {395} (\bibinfo {year} {2002})}\BibitemShut {NoStop}%
\bibitem [{\citenamefont {Sengupta}\ \emph {et~al.}(2003)\citenamefont
  {Sengupta}, \citenamefont {Sandvik},\ and\ \citenamefont
  {Singh}}]{SSS:PRB03}%
  \BibitemOpen
  \bibfield  {author} {\bibinfo {author} {\bibfnamefont {P.}~\bibnamefont
  {Sengupta}}, \bibinfo {author} {\bibfnamefont {A.~W.}\ \bibnamefont
  {Sandvik}}, \ and\ \bibinfo {author} {\bibfnamefont {R.~R.~P.}\ \bibnamefont
  {Singh}},\ }\enquote {\bibinfo {title} {Specific heat of
  quasi-two-dimensional antiferromagnetic Heisenberg models with varying
  interplanar couplings},}\ \href {\doibase 10.1103/PhysRevB.68.094423}
  {\bibfield  {journal} {\bibinfo  {journal} {Phys. Rev. B}\ }\textbf {\bibinfo
  {volume} {68}},\ \bibinfo {pages} {094423} (\bibinfo {year}
  {2003})}\BibitemShut {NoStop}%
\bibitem [{\citenamefont {Schmalfu\ss{}}\ \emph {et~al.}(2005)\citenamefont
  {Schmalfu\ss{}}, \citenamefont {Richter},\ and\ \citenamefont
  {Ihle}}]{SRI:PRB05}%
  \BibitemOpen
  \bibfield  {author} {\bibinfo {author} {\bibfnamefont {D.}~\bibnamefont
  {Schmalfu\ss{}}}, \bibinfo {author} {\bibfnamefont {J.}~\bibnamefont
  {Richter}}, \ and\ \bibinfo {author} {\bibfnamefont {D.}~\bibnamefont
  {Ihle}},\ }\enquote {\bibinfo {title} {Green's function theory of
  quasi-two-dimensional spin-half Heisenberg ferromagnets: Stacked square
  versus stacked kagom\'e lattices},}\ \href {\doibase
  10.1103/PhysRevB.72.224405} {\bibfield  {journal} {\bibinfo  {journal} {Phys.
  Rev. B}\ }\textbf {\bibinfo {volume} {72}},\ \bibinfo {pages} {224405}
  (\bibinfo {year} {2005})}\BibitemShut {NoStop}%
\bibitem [{\citenamefont {Yasuda}\ \emph {et~al.}(2005)\citenamefont {Yasuda},
  \citenamefont {Todo}, \citenamefont {Hukushima}, \citenamefont {Alet},
  \citenamefont {Keller}, \citenamefont {Troyer},\ and\ \citenamefont
  {Takayama}}]{YTH:PRL05}%
  \BibitemOpen
  \bibfield  {author} {\bibinfo {author} {\bibfnamefont {C.}~\bibnamefont
  {Yasuda}}, \bibinfo {author} {\bibfnamefont {S.}~\bibnamefont {Todo}},
  \bibinfo {author} {\bibfnamefont {K.}~\bibnamefont {Hukushima}}, \bibinfo
  {author} {\bibfnamefont {F.}~\bibnamefont {Alet}}, \bibinfo {author}
  {\bibfnamefont {M.}~\bibnamefont {Keller}}, \bibinfo {author} {\bibfnamefont
  {M.}~\bibnamefont {Troyer}}, \ and\ \bibinfo {author} {\bibfnamefont
  {H.}~\bibnamefont {Takayama}},\ }\enquote {\bibinfo {title} {N\'eel
  Temperature of Quasi-Low-Dimensional Heisenberg Antiferromagnets},}\ \href
  {\doibase 10.1103/PhysRevLett.94.217201} {\bibfield  {journal} {\bibinfo
  {journal} {Phys. Rev. Lett.}\ }\textbf {\bibinfo {volume} {94}},\ \bibinfo
  {pages} {217201} (\bibinfo {year} {2005})}\BibitemShut {NoStop}%
\bibitem [{\citenamefont {Lecren}\ \emph {et~al.}(2007)\citenamefont {Lecren},
  \citenamefont {Wernsdorfer}, \citenamefont {Li}, \citenamefont {Vindigni},
  \citenamefont {Miyasaka},\ and\ \citenamefont {Clerac}}]{LWL:JACS07}%
  \BibitemOpen
  \bibfield  {author} {\bibinfo {author} {\bibfnamefont {L.}~\bibnamefont
  {Lecren}}, \bibinfo {author} {\bibfnamefont {W.}~\bibnamefont {Wernsdorfer}},
  \bibinfo {author} {\bibfnamefont {Y.-G.}\ \bibnamefont {Li}}, \bibinfo
  {author} {\bibfnamefont {A.}~\bibnamefont {Vindigni}}, \bibinfo {author}
  {\bibfnamefont {H.}~\bibnamefont {Miyasaka}}, \ and\ \bibinfo {author}
  {\bibfnamefont {R.}~\bibnamefont {Clerac}},\ }\enquote {\bibinfo {title}
  {One-Dimensional Supramolecular Organization of Single-Molecule Magnets},}\
  \href {\doibase 10.1021/ja067744i} {\bibfield  {journal} {\bibinfo  {journal}
  {J. Am. Chem. Soc.}\ }\textbf {\bibinfo {volume} {129}},\ \bibinfo {pages}
  {5045} (\bibinfo {year} {2007})}\BibitemShut {NoStop}%
\bibitem [{\citenamefont {Nie}\ \emph {et~al.}(2008)\citenamefont {Nie},
  \citenamefont {Demeshko}, \citenamefont {Fuchs}, \citenamefont {Dechert},
  \citenamefont {Pruschke},\ and\ \citenamefont {Meyer}}]{NDF:DT08}%
  \BibitemOpen
  \bibfield  {author} {\bibinfo {author} {\bibfnamefont {F.-M.}\ \bibnamefont
  {Nie}}, \bibinfo {author} {\bibfnamefont {S.}~\bibnamefont {Demeshko}},
  \bibinfo {author} {\bibfnamefont {S.}~\bibnamefont {Fuchs}}, \bibinfo
  {author} {\bibfnamefont {S.}~\bibnamefont {Dechert}}, \bibinfo {author}
  {\bibfnamefont {T.}~\bibnamefont {Pruschke}}, \ and\ \bibinfo {author}
  {\bibfnamefont {F.}~\bibnamefont {Meyer}},\ }\enquote {\bibinfo {title}
  {Targeted self-assembly and quantum Monte Carlo magnetic study of an
  alternating nickel(ii) 1D coordination polymer composed of highly
  preorganized binuclear tectons},}\ \href {\doibase 10.1039/B804292F}
  {\bibfield  {journal} {\bibinfo  {journal} {Dalton Trans.}\ ,\ \bibinfo
  {pages} {3971}} (\bibinfo {year} {2008})}\BibitemShut {NoStop}%
\bibitem [{\citenamefont {Butcher}\ \emph {et~al.}(2008)\citenamefont
  {Butcher}, \citenamefont {Landee}, \citenamefont {Turnbull},\ and\
  \citenamefont {Xiao}}]{BLT:ICA08}%
  \BibitemOpen
  \bibfield  {author} {\bibinfo {author} {\bibfnamefont {R.~T.}\ \bibnamefont
  {Butcher}}, \bibinfo {author} {\bibfnamefont {C.~P.}\ \bibnamefont {Landee}},
  \bibinfo {author} {\bibfnamefont {M.~M.}\ \bibnamefont {Turnbull}}, \ and\
  \bibinfo {author} {\bibfnamefont {F.}~\bibnamefont {Xiao}},\ }\enquote
  {\bibinfo {title} {Rectangular two-dimensional antiferromagnetic systems:
  Analysis of copper(II) pyrazine dibromide and dichloride},}\ \href {\doibase
  http://dx.doi.org/10.1016/j.ica.2008.03.090} {\bibfield  {journal} {\bibinfo
  {journal} {Inorg. Chim. Acta}\ }\textbf {\bibinfo {volume} {361}},\ \bibinfo
  {pages} {3654 } (\bibinfo {year} {2008})}\BibitemShut {NoStop}%
\bibitem [{\citenamefont {Morimoto}\ \emph {et~al.}(2009)\citenamefont
  {Morimoto}, \citenamefont {Miyasaka}, \citenamefont {Yamashita},\ and\
  \citenamefont {Irie}}]{MMY:JACS09}%
  \BibitemOpen
  \bibfield  {author} {\bibinfo {author} {\bibfnamefont {M.}~\bibnamefont
  {Morimoto}}, \bibinfo {author} {\bibfnamefont {H.}~\bibnamefont {Miyasaka}},
  \bibinfo {author} {\bibfnamefont {M.}~\bibnamefont {Yamashita}}, \ and\
  \bibinfo {author} {\bibfnamefont {M.}~\bibnamefont {Irie}},\ }\enquote
  {\bibinfo {title} {Coordination Assemblies of [Mn4] Single-Molecule Magnets
  Linked by Photochromic Ligands: Photochemical Control of the Magnetic
  Properties},}\ \href {\doibase 10.1021/ja903366d} {\bibfield  {journal}
  {\bibinfo  {journal} {J. Am. Chem. Soc.}\ }\textbf {\bibinfo {volume}
  {131}},\ \bibinfo {pages} {9823} (\bibinfo {year} {2009})}\BibitemShut
  {NoStop}%
\bibitem [{\citenamefont {Arango}\ \emph {et~al.}(2011)\citenamefont {Arango},
  \citenamefont {Vavilova}, \citenamefont {Abdel-Hafiez}, \citenamefont
  {Janson}, \citenamefont {Tsirlin}, \citenamefont {Rosner}, \citenamefont
  {Drechsler}, \citenamefont {Weil}, \citenamefont {N\'enert}, \citenamefont
  {Klingeler}, \citenamefont {Volkova}, \citenamefont {Vasiliev}, \citenamefont
  {Kataev},\ and\ \citenamefont {B\"uchner}}]{AVA:PRB11}%
  \BibitemOpen
  \bibfield  {author} {\bibinfo {author} {\bibfnamefont {Y.~C.}\ \bibnamefont
  {Arango}}, \bibinfo {author} {\bibfnamefont {E.}~\bibnamefont {Vavilova}},
  \bibinfo {author} {\bibfnamefont {M.}~\bibnamefont {Abdel-Hafiez}}, \bibinfo
  {author} {\bibfnamefont {O.}~\bibnamefont {Janson}}, \bibinfo {author}
  {\bibfnamefont {A.~A.}\ \bibnamefont {Tsirlin}}, \bibinfo {author}
  {\bibfnamefont {H.}~\bibnamefont {Rosner}}, \bibinfo {author} {\bibfnamefont
  {S.-L.}\ \bibnamefont {Drechsler}}, \bibinfo {author} {\bibfnamefont
  {M.}~\bibnamefont {Weil}}, \bibinfo {author} {\bibfnamefont {G.}~\bibnamefont
  {N\'enert}}, \bibinfo {author} {\bibfnamefont {R.}~\bibnamefont {Klingeler}},
  \bibinfo {author} {\bibfnamefont {O.}~\bibnamefont {Volkova}}, \bibinfo
  {author} {\bibfnamefont {A.}~\bibnamefont {Vasiliev}}, \bibinfo {author}
  {\bibfnamefont {V.}~\bibnamefont {Kataev}}, \ and\ \bibinfo {author}
  {\bibfnamefont {B.}~\bibnamefont {B\"uchner}},\ }\enquote {\bibinfo {title}
  {Magnetic properties of the low-dimensional spin-$\frac{1}{2}$ magnet
  $\ensuremath{\alpha}$-Cu${}_{2}$As${}_{2}$O${}_{7}$},}\ \href {\doibase
  10.1103/PhysRevB.84.134430} {\bibfield  {journal} {\bibinfo  {journal} {Phys.
  Rev. B}\ }\textbf {\bibinfo {volume} {84}},\ \bibinfo {pages} {134430}
  (\bibinfo {year} {2011})}\BibitemShut {NoStop}%
\bibitem [{\citenamefont {Shapira}\ and\ \citenamefont
  {Bindilatti}(2002)}]{ShB:JAP02}%
  \BibitemOpen
  \bibfield  {author} {\bibinfo {author} {\bibfnamefont {Y.}~\bibnamefont
  {Shapira}}\ and\ \bibinfo {author} {\bibfnamefont {V.}~\bibnamefont
  {Bindilatti}},\ }\enquote {\bibinfo {title} {Magnetization-step studies of
  antiferromagnetic clusters and single ions: Exchange, anisotropy, and
  statistics},}\ \href {\doibase 10.1063/1.1507808} {\bibfield  {journal}
  {\bibinfo  {journal} {J. Appl. Phys.}\ }\textbf {\bibinfo {volume} {92}},\
  \bibinfo {pages} {4155} (\bibinfo {year} {2002})}\BibitemShut {NoStop}%
\bibitem [{\citenamefont {Schr\"{o}der}\ \emph {et~al.}(2008)\citenamefont
  {Schr\"{o}der}, \citenamefont {Prozorov}, \citenamefont {K\"{o}gerler},
  \citenamefont {Vannette}, \citenamefont {Fang}, \citenamefont {Luban},
  \citenamefont {Matsuo}, \citenamefont {Kindo}, \citenamefont {M\"{u}ller},\
  and\ \citenamefont {Todea}}]{SPK:PRB08}%
  \BibitemOpen
  \bibfield  {author} {\bibinfo {author} {\bibfnamefont {C.}~\bibnamefont
  {Schr\"{o}der}}, \bibinfo {author} {\bibfnamefont {R.}~\bibnamefont
  {Prozorov}}, \bibinfo {author} {\bibfnamefont {P.}~\bibnamefont
  {K\"{o}gerler}}, \bibinfo {author} {\bibfnamefont {M.~D.}\ \bibnamefont
  {Vannette}}, \bibinfo {author} {\bibfnamefont {X.}~\bibnamefont {Fang}},
  \bibinfo {author} {\bibfnamefont {M.}~\bibnamefont {Luban}}, \bibinfo
  {author} {\bibfnamefont {A.}~\bibnamefont {Matsuo}}, \bibinfo {author}
  {\bibfnamefont {K.}~\bibnamefont {Kindo}}, \bibinfo {author} {\bibfnamefont
  {A.}~\bibnamefont {M\"{u}ller}}, \ and\ \bibinfo {author} {\bibfnamefont
  {A.~M.}\ \bibnamefont {Todea}},\ }\enquote {\bibinfo {title} {Multiple
  nearest-neighbor exchange model for the frustrated magnetic molecules
  \{Mo$_{72}$Fe$_{30}$\} and \{Mo$_{72}$Cr$_{30}$\}},}\ \href {\doibase
  10.1103/PhysRevB.77.224409} {\bibfield  {journal} {\bibinfo  {journal} {Phys.
  Rev. B}\ }\textbf {\bibinfo {volume} {77}},\ \bibinfo {pages} {224409}
  (\bibinfo {year} {2008})}\BibitemShut {NoStop}%
\bibitem [{\citenamefont {K{\"u}hne}\ \emph {et~al.}(2015)\citenamefont
  {K{\"u}hne}, \citenamefont {Kostakis}, \citenamefont {Anson},\ and\
  \citenamefont {Powell}}]{KKA:CC15}%
  \BibitemOpen
  \bibfield  {author} {\bibinfo {author} {\bibfnamefont {I.~A.}\ \bibnamefont
  {K{\"u}hne}}, \bibinfo {author} {\bibfnamefont {G.~E.}\ \bibnamefont
  {Kostakis}}, \bibinfo {author} {\bibfnamefont {C.~E.}\ \bibnamefont {Anson}},
  \ and\ \bibinfo {author} {\bibfnamefont {A.~K.}\ \bibnamefont {Powell}},\
  }\enquote {\bibinfo {title} {A magnetically highly frustrated
  Cu$^{\text{II}}_{27}$ coordination cluster containing a Cu$_{18}$
  folded-sheet motif},}\ \href {\doibase 10.1039/C4CC09469G} {\bibfield
  {journal} {\bibinfo  {journal} {Chem. Commun.}\ }\textbf {\bibinfo {volume}
  {51}},\ \bibinfo {pages} {2702} (\bibinfo {year} {2015})}\BibitemShut
  {NoStop}%
\bibitem [{\citenamefont {Palacios}\ \emph {et~al.}(2015)\citenamefont
  {Palacios}, \citenamefont {Pineda}, \citenamefont {Sanz}, \citenamefont
  {Inglis}, \citenamefont {Pitak}, \citenamefont {Coles}, \citenamefont
  {Evangelisti}, \citenamefont {Nojiri}, \citenamefont {Heesing}, \citenamefont
  {Brechin}, \citenamefont {Schnack},\ and\ \citenamefont
  {Winpenny}}]{PPS:CPC15}%
  \BibitemOpen
  \bibfield  {author} {\bibinfo {author} {\bibfnamefont {M.~A.}\ \bibnamefont
  {Palacios}}, \bibinfo {author} {\bibfnamefont {E.~M.}\ \bibnamefont
  {Pineda}}, \bibinfo {author} {\bibfnamefont {S.}~\bibnamefont {Sanz}},
  \bibinfo {author} {\bibfnamefont {R.}~\bibnamefont {Inglis}}, \bibinfo
  {author} {\bibfnamefont {M.~B.}\ \bibnamefont {Pitak}}, \bibinfo {author}
  {\bibfnamefont {S.~J.}\ \bibnamefont {Coles}}, \bibinfo {author}
  {\bibfnamefont {M.}~\bibnamefont {Evangelisti}}, \bibinfo {author}
  {\bibfnamefont {H.}~\bibnamefont {Nojiri}}, \bibinfo {author} {\bibfnamefont
  {C.}~\bibnamefont {Heesing}}, \bibinfo {author} {\bibfnamefont {E.~K.}\
  \bibnamefont {Brechin}}, \bibinfo {author} {\bibfnamefont {J.}~\bibnamefont
  {Schnack}}, \ and\ \bibinfo {author} {\bibfnamefont {R.~E.~P.}\ \bibnamefont
  {Winpenny}},\ }\enquote {\bibinfo {title} {Copper Keplerates: High Symmetry
  Magnetic Molecules},}\ \href {\doibase 10.1002/cphc.201500956} {\bibfield
  {journal} {\bibinfo  {journal} {ChemPhysChem}\ } (\bibinfo {year} {2015}),\
  10.1002/cphc.201500956},\ \bibinfo {note} {in print}\BibitemShut {NoStop}%
\bibitem [{\citenamefont {Delfs}\ \emph {et~al.}(1993)\citenamefont {Delfs},
  \citenamefont {Gatteschi}, \citenamefont {Pardi}, \citenamefont {Sessoli},
  \citenamefont {Wieghardt},\ and\ \citenamefont {Hanke}}]{DGP:IC93}%
  \BibitemOpen
  \bibfield  {author} {\bibinfo {author} {\bibfnamefont {C.}~\bibnamefont
  {Delfs}}, \bibinfo {author} {\bibfnamefont {D.}~\bibnamefont {Gatteschi}},
  \bibinfo {author} {\bibfnamefont {L.}~\bibnamefont {Pardi}}, \bibinfo
  {author} {\bibfnamefont {R.}~\bibnamefont {Sessoli}}, \bibinfo {author}
  {\bibfnamefont {K.}~\bibnamefont {Wieghardt}}, \ and\ \bibinfo {author}
  {\bibfnamefont {D.}~\bibnamefont {Hanke}},\ }\enquote {\bibinfo {title}
  {Magnetic properties of an octanuclear iron(III) cation},}\ \href {\doibase
  10.1021/ic00066a022} {\bibfield  {journal} {\bibinfo  {journal} {Inorg.
  Chem.}\ }\textbf {\bibinfo {volume} {32}},\ \bibinfo {pages} {3099} (\bibinfo
  {year} {1993})}\BibitemShut {NoStop}%
\bibitem [{\citenamefont {Gatteschi}\ and\ \citenamefont
  {Pardi}(1993)}]{GaP:GCI93}%
  \BibitemOpen
  \bibfield  {author} {\bibinfo {author} {\bibfnamefont {D.}~\bibnamefont
  {Gatteschi}}\ and\ \bibinfo {author} {\bibfnamefont {L.}~\bibnamefont
  {Pardi}},\ }\enquote {\bibinfo {title} {Magnetic-properties of
  high-nuclearity spin clusters - a fast and efficient procedure for the
  calculation of the energy-levels},}\ \href@noop {} {\bibfield  {journal}
  {\bibinfo  {journal} {Gazz. Chim. Ital.}\ }\textbf {\bibinfo {volume}
  {123}},\ \bibinfo {pages} {231} (\bibinfo {year} {1993})}\BibitemShut
  {NoStop}%
\bibitem [{\citenamefont {Borras-Almenar}\ \emph {et~al.}(1999)\citenamefont
  {Borras-Almenar}, \citenamefont {Clemente-Juan}, \citenamefont {Coronado},\
  and\ \citenamefont {Tsukerblat}}]{BCC:IC99}%
  \BibitemOpen
  \bibfield  {author} {\bibinfo {author} {\bibfnamefont {J.~J.}\ \bibnamefont
  {Borras-Almenar}}, \bibinfo {author} {\bibfnamefont {J.~M.}\ \bibnamefont
  {Clemente-Juan}}, \bibinfo {author} {\bibfnamefont {E.}~\bibnamefont
  {Coronado}}, \ and\ \bibinfo {author} {\bibfnamefont {B.~S.}\ \bibnamefont
  {Tsukerblat}},\ }\enquote {\bibinfo {title} {High-nuclearity magnetic
  clusters: Generalized spin Hamiltonian and its use for the calculation of the
  energy levels, bulk magnetic properties, and inelastic neutron scattering
  spectra},}\ \href {http://dx.doi.org/10.1021/ic990915i} {\bibfield  {journal}
  {\bibinfo  {journal} {Inorg. Chem.}\ }\textbf {\bibinfo {volume} {38}},\
  \bibinfo {pages} {6081} (\bibinfo {year} {1999})}\BibitemShut {NoStop}%
\bibitem [{\citenamefont {Bencini}\ and\ \citenamefont
  {Gatteschi}(1990)}]{BeG:EPR}%
  \BibitemOpen
  \bibfield  {author} {\bibinfo {author} {\bibfnamefont {A.}~\bibnamefont
  {Bencini}}\ and\ \bibinfo {author} {\bibfnamefont {D.}~\bibnamefont
  {Gatteschi}},\ }\href@noop {} {\emph {\bibinfo {title} {Electron paramagnetic
  resonance of exchange coupled systems}}}\ (\bibinfo  {publisher} {Springer},\
  \bibinfo {address} {Berlin, Heidelberg},\ \bibinfo {year} {1990})\BibitemShut
  {NoStop}%
\bibitem [{\citenamefont {Tsukerblat}(2006)}]{Tsu:group_theory}%
  \BibitemOpen
  \bibfield  {author} {\bibinfo {author} {\bibfnamefont {B.~S.}\ \bibnamefont
  {Tsukerblat}},\ }\href@noop {} {\emph {\bibinfo {title} {Group theory in
  chemistry and spectroscopy: a simple guide to advanced usage}}},\ \bibinfo
  {edition} {2nd}\ ed.\ (\bibinfo  {publisher} {Dover Publications},\ \bibinfo
  {address} {Mineola, New York},\ \bibinfo {year} {2006})\BibitemShut {NoStop}%
\bibitem [{\citenamefont {Waldmann}(2000)}]{Wal:PRB00}%
  \BibitemOpen
  \bibfield  {author} {\bibinfo {author} {\bibfnamefont {O.}~\bibnamefont
  {Waldmann}},\ }\enquote {\bibinfo {title} {Symmetry and energy spectrum of
  high-nuclearity spin clusters},}\ \href {\doibase 10.1103/PhysRevB.61.6138}
  {\bibfield  {journal} {\bibinfo  {journal} {Phys. Rev. B}\ }\textbf {\bibinfo
  {volume} {61}},\ \bibinfo {pages} {6138} (\bibinfo {year}
  {2000})}\BibitemShut {NoStop}%
\bibitem [{\citenamefont {Tsukerblat}(2008)}]{Tsu:ICA08}%
  \BibitemOpen
  \bibfield  {author} {\bibinfo {author} {\bibfnamefont {B.}~\bibnamefont
  {Tsukerblat}},\ }\enquote {\bibinfo {title} {Group-theoretical approaches in
  molecular magnetism: Metal clusters},}\ \href {\doibase
  10.1016/j.ica.2008.03.012} {\bibfield  {journal} {\bibinfo  {journal} {Inorg.
  Chim. Acta}\ }\textbf {\bibinfo {volume} {361}},\ \bibinfo {pages} {3746}
  (\bibinfo {year} {2008})}\BibitemShut {NoStop}%
\bibitem [{\citenamefont {Boyarchenkov}\ \emph {et~al.}(2007)\citenamefont
  {Boyarchenkov}, \citenamefont {Bostrem},\ and\ \citenamefont
  {Ovchinnikov}}]{BBO:PRB07}%
  \BibitemOpen
  \bibfield  {author} {\bibinfo {author} {\bibfnamefont {A.~S.}\ \bibnamefont
  {Boyarchenkov}}, \bibinfo {author} {\bibfnamefont {I.~G.}\ \bibnamefont
  {Bostrem}}, \ and\ \bibinfo {author} {\bibfnamefont {A.~S.}\ \bibnamefont
  {Ovchinnikov}},\ }\enquote {\bibinfo {title} {Quantum magnetization plateau
  and sign change of the magnetocaloric effect in a ferrimagnetic spin
  chain.}}\ \href {http://link.aps.org/abstract/PRB/v76/e224410} {\bibfield
  {journal} {\bibinfo  {journal} {Phys. Rev. B}\ }\textbf {\bibinfo {volume}
  {76}},\ \bibinfo {pages} {224410} (\bibinfo {year} {2007})}\BibitemShut
  {NoStop}%
\bibitem [{\citenamefont {Schnalle}\ and\ \citenamefont
  {Schnack}(2009)}]{ScS:PRB09}%
  \BibitemOpen
  \bibfield  {author} {\bibinfo {author} {\bibfnamefont {R.}~\bibnamefont
  {Schnalle}}\ and\ \bibinfo {author} {\bibfnamefont {J.}~\bibnamefont
  {Schnack}},\ }\enquote {\bibinfo {title} {Numerically exact and approximate
  determination of energy eigenvalues for antiferromagnetic molecules using
  irreducible tensor operators and general point-group symmetries},}\ \href
  {http://link.aps.org/abstract/PRB/v79/e104419} {\bibfield  {journal}
  {\bibinfo  {journal} {Phys. Rev. B}\ }\textbf {\bibinfo {volume} {79}},\
  \bibinfo {pages} {104419} (\bibinfo {year} {2009})}\BibitemShut {NoStop}%
\bibitem [{\citenamefont {Schnalle}\ and\ \citenamefont
  {Schnack}(2010)}]{ScS:IRPC10}%
  \BibitemOpen
  \bibfield  {author} {\bibinfo {author} {\bibfnamefont {R.}~\bibnamefont
  {Schnalle}}\ and\ \bibinfo {author} {\bibfnamefont {J.}~\bibnamefont
  {Schnack}},\ }\enquote {\bibinfo {title} {Calculating the energy spectra of
  magnetic molecules: application of real- and spin-space symmetries},}\ \href
  {http://dx.doi.org/10.1080/0144235X.2010.485755} {\bibfield  {journal}
  {\bibinfo  {journal} {Int. Rev. Phys. Chem.}\ }\textbf {\bibinfo {volume}
  {29}},\ \bibinfo {pages} {403} (\bibinfo {year} {2010})}\BibitemShut
  {NoStop}%
\bibitem [{\citenamefont {Sandvik}\ and\ \citenamefont
  {Kurkij\"arvi}(1991)}]{SaK:PRB91}%
  \BibitemOpen
  \bibfield  {author} {\bibinfo {author} {\bibfnamefont {A.~W.}\ \bibnamefont
  {Sandvik}}\ and\ \bibinfo {author} {\bibfnamefont {J.}~\bibnamefont
  {Kurkij\"arvi}},\ }\enquote {\bibinfo {title} {Quantum Monte Carlo simulation
  method for spin systems},}\ \href {\doibase 10.1103/PhysRevB.43.5950}
  {\bibfield  {journal} {\bibinfo  {journal} {Phys. Rev. B}\ }\textbf {\bibinfo
  {volume} {43}},\ \bibinfo {pages} {5950} (\bibinfo {year}
  {1991})}\BibitemShut {NoStop}%
\bibitem [{\citenamefont {Sandvik}(1999)}]{San:PRB99}%
  \BibitemOpen
  \bibfield  {author} {\bibinfo {author} {\bibfnamefont {A.~W.}\ \bibnamefont
  {Sandvik}},\ }\enquote {\bibinfo {title} {Stochastic series expansion method
  with operator-loop update},}\ \href {\doibase 10.1103/PhysRevB.59.R14157}
  {\bibfield  {journal} {\bibinfo  {journal} {Phys. Rev. B}\ }\textbf {\bibinfo
  {volume} {59}},\ \bibinfo {pages} {R14157} (\bibinfo {year}
  {1999})}\BibitemShut {NoStop}%
\bibitem [{\citenamefont {Sylju\aa{}sen}\ and\ \citenamefont
  {Sandvik}(2002)}]{SyS:PRE02}%
  \BibitemOpen
  \bibfield  {author} {\bibinfo {author} {\bibfnamefont {O.~F.}\ \bibnamefont
  {Sylju\aa{}sen}}\ and\ \bibinfo {author} {\bibfnamefont {A.~W.}\ \bibnamefont
  {Sandvik}},\ }\enquote {\bibinfo {title} {Quantum Monte Carlo with directed
  loops},}\ \href {\doibase 10.1103/PhysRevE.66.046701} {\bibfield  {journal}
  {\bibinfo  {journal} {Phys. Rev. E}\ }\textbf {\bibinfo {volume} {66}},\
  \bibinfo {pages} {046701} (\bibinfo {year} {2002})}\BibitemShut {NoStop}%
\bibitem [{\citenamefont {Schulz}(1996)}]{Sch:PRL96}%
  \BibitemOpen
  \bibfield  {author} {\bibinfo {author} {\bibfnamefont {H.~J.}\ \bibnamefont
  {Schulz}},\ }\enquote {\bibinfo {title} {Dynamics of Coupled Quantum Spin
  Chains},}\ \href {\doibase 10.1103/PhysRevLett.77.2790} {\bibfield  {journal}
  {\bibinfo  {journal} {Phys. Rev. Lett.}\ }\textbf {\bibinfo {volume} {77}},\
  \bibinfo {pages} {2790} (\bibinfo {year} {1996})}\BibitemShut {NoStop}%
\bibitem [{\citenamefont {Laflorencie}\ \emph {et~al.}(2006)\citenamefont
  {Laflorencie}, \citenamefont {Wessel}, \citenamefont {L\"auchli},\ and\
  \citenamefont {Rieger}}]{LWL:PRB06}%
  \BibitemOpen
  \bibfield  {author} {\bibinfo {author} {\bibfnamefont {N.}~\bibnamefont
  {Laflorencie}}, \bibinfo {author} {\bibfnamefont {S.}~\bibnamefont {Wessel}},
  \bibinfo {author} {\bibfnamefont {A.}~\bibnamefont {L\"auchli}}, \ and\
  \bibinfo {author} {\bibfnamefont {H.}~\bibnamefont {Rieger}},\ }\enquote
  {\bibinfo {title} {Random-exchange quantum Heisenberg antiferromagnets on a
  square lattice},}\ \href {\doibase 10.1103/PhysRevB.73.060403} {\bibfield
  {journal} {\bibinfo  {journal} {Phys. Rev. B}\ }\textbf {\bibinfo {volume}
  {73}},\ \bibinfo {pages} {060403} (\bibinfo {year} {2006})}\BibitemShut
  {NoStop}%
\bibitem [{\citenamefont {Richter}\ \emph {et~al.}(1996)\citenamefont
  {Richter}, \citenamefont {Voigt}, \citenamefont {Kr{\"u}ger},\ and\
  \citenamefont {Gros}}]{RVK:JPAMG96}%
  \BibitemOpen
  \bibfield  {author} {\bibinfo {author} {\bibfnamefont {J.}~\bibnamefont
  {Richter}}, \bibinfo {author} {\bibfnamefont {A.}~\bibnamefont {Voigt}},
  \bibinfo {author} {\bibfnamefont {S.~E.}\ \bibnamefont {Kr{\"u}ger}}, \ and\
  \bibinfo {author} {\bibfnamefont {C.}~\bibnamefont {Gros}},\ }\enquote
  {\bibinfo {title} {The spin-1/2 Heisenberg star with frustration: II. The
  influence of the embedding medium},}\ \href
  {http://stacks.iop.org/0305-4470/29/i=4/a=010} {\bibfield  {journal}
  {\bibinfo  {journal} {Journal of Physics A: Mathematical and General}\
  }\textbf {\bibinfo {volume} {29}},\ \bibinfo {pages} {825} (\bibinfo {year}
  {1996})}\BibitemShut {NoStop}%
\bibitem [{\citenamefont {Albuquerque}\ \emph {et~al.}(2007)\citenamefont
  {Albuquerque}, \citenamefont {Alet}, \citenamefont {Corboz}, \citenamefont
  {Dayal}, \citenamefont {Feiguin}, \citenamefont {Fuchs}, \citenamefont
  {Gamper}, \citenamefont {Gull}, \citenamefont {G{\"u}rtler}, \citenamefont
  {Honecker}, \citenamefont {Igarashi}, \citenamefont {K{\"o}rner},
  \citenamefont {Kozhevnikov}, \citenamefont {L{\"a}uchli}, \citenamefont
  {Manmana}, \citenamefont {Matsumoto}, \citenamefont {McCulloch},
  \citenamefont {Michel}, \citenamefont {Noack}, \citenamefont {Pawlowski},
  \citenamefont {Pollet}, \citenamefont {Pruschke}, \citenamefont
  {Schollw{\"o}ck}, \citenamefont {Todo}, \citenamefont {Trebst}, \citenamefont
  {Troyer}, \citenamefont {Werner},\ and\ \citenamefont
  {Wessel}}]{ALPS:JMMM07}%
  \BibitemOpen
  \bibfield  {author} {\bibinfo {author} {\bibfnamefont {A.}~\bibnamefont
  {Albuquerque}}, \bibinfo {author} {\bibfnamefont {F.}~\bibnamefont {Alet}},
  \bibinfo {author} {\bibfnamefont {P.}~\bibnamefont {Corboz}}, \bibinfo
  {author} {\bibfnamefont {P.}~\bibnamefont {Dayal}}, \bibinfo {author}
  {\bibfnamefont {A.}~\bibnamefont {Feiguin}}, \bibinfo {author} {\bibfnamefont
  {S.}~\bibnamefont {Fuchs}}, \bibinfo {author} {\bibfnamefont
  {L.}~\bibnamefont {Gamper}}, \bibinfo {author} {\bibfnamefont
  {E.}~\bibnamefont {Gull}}, \bibinfo {author} {\bibfnamefont {S.}~\bibnamefont
  {G{\"u}rtler}}, \bibinfo {author} {\bibfnamefont {A.}~\bibnamefont
  {Honecker}}, \bibinfo {author} {\bibfnamefont {R.}~\bibnamefont {Igarashi}},
  \bibinfo {author} {\bibfnamefont {M.}~\bibnamefont {K{\"o}rner}}, \bibinfo
  {author} {\bibfnamefont {A.}~\bibnamefont {Kozhevnikov}}, \bibinfo {author}
  {\bibfnamefont {A.}~\bibnamefont {L{\"a}uchli}}, \bibinfo {author}
  {\bibfnamefont {S.}~\bibnamefont {Manmana}}, \bibinfo {author} {\bibfnamefont
  {M.}~\bibnamefont {Matsumoto}}, \bibinfo {author} {\bibfnamefont
  {I.}~\bibnamefont {McCulloch}}, \bibinfo {author} {\bibfnamefont
  {F.}~\bibnamefont {Michel}}, \bibinfo {author} {\bibfnamefont
  {R.}~\bibnamefont {Noack}}, \bibinfo {author} {\bibfnamefont
  {G.}~\bibnamefont {Pawlowski}}, \bibinfo {author} {\bibfnamefont
  {L.}~\bibnamefont {Pollet}}, \bibinfo {author} {\bibfnamefont
  {T.}~\bibnamefont {Pruschke}}, \bibinfo {author} {\bibfnamefont
  {U.}~\bibnamefont {Schollw{\"o}ck}}, \bibinfo {author} {\bibfnamefont
  {S.}~\bibnamefont {Todo}}, \bibinfo {author} {\bibfnamefont {S.}~\bibnamefont
  {Trebst}}, \bibinfo {author} {\bibfnamefont {M.}~\bibnamefont {Troyer}},
  \bibinfo {author} {\bibfnamefont {P.}~\bibnamefont {Werner}}, \ and\ \bibinfo
  {author} {\bibfnamefont {S.}~\bibnamefont {Wessel}} (\bibinfo {collaboration}
  {ALPS collaboration}),\ }\enquote {\bibinfo {title} {The ALPS project release
  1.3: Open-source software for strongly correlated systems},}\ \href
  {http://www.sciencedirect.com/science/article/B6TJJ-4M87BSK-1V/2/94142e0f273368ee9c12126c5f9e8c82}
  {\bibfield  {journal} {\bibinfo  {journal} {J. Magn. Magn. Mater.}\ }\textbf
  {\bibinfo {volume} {310}},\ \bibinfo {pages} {1187 } (\bibinfo {year}
  {2007})}\BibitemShut {NoStop}%
\bibitem [{\citenamefont {Bauer}\ \emph {et~al.}(2011)\citenamefont {Bauer},
  \citenamefont {Carr}, \citenamefont {Evertz}, \citenamefont {Feiguin},
  \citenamefont {Freire}, \citenamefont {Fuchs}, \citenamefont {Gamper},
  \citenamefont {Gukelberger}, \citenamefont {Gull}, \citenamefont {Guertler},
  \citenamefont {Hehn}, \citenamefont {Igarashi}, \citenamefont {Isakov},
  \citenamefont {Koop}, \citenamefont {Ma}, \citenamefont {Mates},
  \citenamefont {Matsuo}, \citenamefont {Parcollet}, \citenamefont
  {Pawłowski}, \citenamefont {Picon}, \citenamefont {Pollet}, \citenamefont
  {Santos}, \citenamefont {Scarola}, \citenamefont {Schollw{\"o}ck},
  \citenamefont {Silva}, \citenamefont {Surer}, \citenamefont {Todo},
  \citenamefont {Trebst}, \citenamefont {Troyer}, \citenamefont {Wall},
  \citenamefont {Werner},\ and\ \citenamefont {Wessel}}]{BCE:JSMTE11}%
  \BibitemOpen
  \bibfield  {author} {\bibinfo {author} {\bibfnamefont {B.}~\bibnamefont
  {Bauer}}, \bibinfo {author} {\bibfnamefont {L.~D.}\ \bibnamefont {Carr}},
  \bibinfo {author} {\bibfnamefont {H.~G.}\ \bibnamefont {Evertz}}, \bibinfo
  {author} {\bibfnamefont {A.}~\bibnamefont {Feiguin}}, \bibinfo {author}
  {\bibfnamefont {J.}~\bibnamefont {Freire}}, \bibinfo {author} {\bibfnamefont
  {S.}~\bibnamefont {Fuchs}}, \bibinfo {author} {\bibfnamefont
  {L.}~\bibnamefont {Gamper}}, \bibinfo {author} {\bibfnamefont
  {J.}~\bibnamefont {Gukelberger}}, \bibinfo {author} {\bibfnamefont
  {E.}~\bibnamefont {Gull}}, \bibinfo {author} {\bibfnamefont {S.}~\bibnamefont
  {Guertler}}, \bibinfo {author} {\bibfnamefont {A.}~\bibnamefont {Hehn}},
  \bibinfo {author} {\bibfnamefont {R.}~\bibnamefont {Igarashi}}, \bibinfo
  {author} {\bibfnamefont {S.~V.}\ \bibnamefont {Isakov}}, \bibinfo {author}
  {\bibfnamefont {D.}~\bibnamefont {Koop}}, \bibinfo {author} {\bibfnamefont
  {P.~N.}\ \bibnamefont {Ma}}, \bibinfo {author} {\bibfnamefont
  {P.}~\bibnamefont {Mates}}, \bibinfo {author} {\bibfnamefont
  {H.}~\bibnamefont {Matsuo}}, \bibinfo {author} {\bibfnamefont
  {O.}~\bibnamefont {Parcollet}}, \bibinfo {author} {\bibfnamefont
  {G.}~\bibnamefont {Pawłowski}}, \bibinfo {author} {\bibfnamefont {J.~D.}\
  \bibnamefont {Picon}}, \bibinfo {author} {\bibfnamefont {L.}~\bibnamefont
  {Pollet}}, \bibinfo {author} {\bibfnamefont {E.}~\bibnamefont {Santos}},
  \bibinfo {author} {\bibfnamefont {V.~W.}\ \bibnamefont {Scarola}}, \bibinfo
  {author} {\bibfnamefont {U.}~\bibnamefont {Schollw{\"o}ck}}, \bibinfo
  {author} {\bibfnamefont {C.}~\bibnamefont {Silva}}, \bibinfo {author}
  {\bibfnamefont {B.}~\bibnamefont {Surer}}, \bibinfo {author} {\bibfnamefont
  {S.}~\bibnamefont {Todo}}, \bibinfo {author} {\bibfnamefont {S.}~\bibnamefont
  {Trebst}}, \bibinfo {author} {\bibfnamefont {M.}~\bibnamefont {Troyer}},
  \bibinfo {author} {\bibfnamefont {M.~L.}\ \bibnamefont {Wall}}, \bibinfo
  {author} {\bibfnamefont {P.}~\bibnamefont {Werner}}, \ and\ \bibinfo {author}
  {\bibfnamefont {S.}~\bibnamefont {Wessel}},\ }\enquote {\bibinfo {title} {The
  ALPS project release 2.0: open source software for strongly correlated
  systems},}\ \href {http://stacks.iop.org/1742-5468/2011/i=05/a=P05001}
  {\bibfield  {journal} {\bibinfo  {journal} {J. Stat. Mech.: Theor. Exp.}\
  }\textbf {\bibinfo {volume} {2011}},\ \bibinfo {pages} {P05001} (\bibinfo
  {year} {2011})}\BibitemShut {NoStop}%
\bibitem [{\citenamefont {Bethe}(1931)}]{Bet:ZP31}%
  \BibitemOpen
  \bibfield  {author} {\bibinfo {author} {\bibfnamefont {H.}~\bibnamefont
  {Bethe}},\ }\enquote {\bibinfo {title} {Zur Theorie der Metalle},}\
  \href@noop {} {\bibfield  {journal} {\bibinfo  {journal} {Z. Phys.}\ }\textbf
  {\bibinfo {volume} {71}},\ \bibinfo {pages} {205} (\bibinfo {year}
  {1931})}\BibitemShut {NoStop}%
\bibitem [{\citenamefont {Hulth{\'e}n}(1938)}]{Hul:AMAFA38}%
  \BibitemOpen
  \bibfield  {author} {\bibinfo {author} {\bibfnamefont {L.}~\bibnamefont
  {Hulth{\'e}n}},\ }\enquote {\bibinfo {title} {{\"U}ber das Austauschproblem
  eines Kristalles},}\ \href@noop {} {\bibfield  {journal} {\bibinfo  {journal}
  {Arkiv Mat. Astr. Fys. A}\ }\textbf {\bibinfo {volume} {26}},\ \bibinfo
  {pages} {106} (\bibinfo {year} {1938})}\BibitemShut {NoStop}%
\bibitem [{\citenamefont {Griffiths}(1964)}]{Gri:PR64}%
  \BibitemOpen
  \bibfield  {author} {\bibinfo {author} {\bibfnamefont {R.~B.}\ \bibnamefont
  {Griffiths}},\ }\enquote {\bibinfo {title} {Magnetization Curve at Zero
  Temperature for the Antiferromagnetic Heisenberg Linear Chain},}\ \href
  {\doibase 10.1103/PhysRev.133.A768} {\bibfield  {journal} {\bibinfo
  {journal} {Phys. Rev.}\ }\textbf {\bibinfo {volume} {133}},\ \bibinfo {pages}
  {A768} (\bibinfo {year} {1964})}\BibitemShut {NoStop}%
\bibitem [{\citenamefont {Haldane}(1983{\natexlab{a}})}]{Hal:PL83}%
  \BibitemOpen
  \bibfield  {author} {\bibinfo {author} {\bibfnamefont {F.}~\bibnamefont
  {Haldane}},\ }\enquote {\bibinfo {title} {Continuum dynamics of the 1-d
  {H}eisenberg anti-ferromagnet - identification with the o(3) non-linear
  sigma-model},}\ \href {\doibase 10.1016/0375-9601(83)90631-X} {\bibfield
  {journal} {\bibinfo  {journal} {Phys. Lett. A}\ }\textbf {\bibinfo {volume}
  {93}},\ \bibinfo {pages} {464} (\bibinfo {year}
  {1983}{\natexlab{a}})}\BibitemShut {NoStop}%
\bibitem [{\citenamefont {Haldane}(1983{\natexlab{b}})}]{Hal:PRL83}%
  \BibitemOpen
  \bibfield  {author} {\bibinfo {author} {\bibfnamefont {F.}~\bibnamefont
  {Haldane}},\ }\enquote {\bibinfo {title} {Non-linear field-theory of
  large-spin Heisenberg antiferromagnets - semi-classically quantized solitons
  of the onedimensional easy-axis neel state},}\ \href {\doibase
  10.1103/PhysRevLett.50.1153} {\bibfield  {journal} {\bibinfo  {journal}
  {Phys. Rev. Lett.}\ }\textbf {\bibinfo {volume} {50}},\ \bibinfo {pages}
  {1153} (\bibinfo {year} {1983}{\natexlab{b}})}\BibitemShut {NoStop}%
\bibitem [{\citenamefont {Affleck}\ and\ \citenamefont
  {Lieb}(1986)}]{AfL:LMP86}%
  \BibitemOpen
  \bibfield  {author} {\bibinfo {author} {\bibfnamefont {I.}~\bibnamefont
  {Affleck}}\ and\ \bibinfo {author} {\bibfnamefont {E.~H.}\ \bibnamefont
  {Lieb}},\ }\enquote {\bibinfo {title} {A proof of part of Haldane's
  conjecture on spin chains},}\ \href {\doibase 10.1007/BF00400304} {\bibfield
  {journal} {\bibinfo  {journal} {Lett. Math. Phys.}\ }\textbf {\bibinfo
  {volume} {12}},\ \bibinfo {pages} {57} (\bibinfo {year} {1986})}\BibitemShut
  {NoStop}%
\bibitem [{\citenamefont {Affleck}\ \emph {et~al.}(1989)\citenamefont
  {Affleck}, \citenamefont {Gepner}, \citenamefont {Schulz},\ and\
  \citenamefont {Ziman}}]{AGS:JPA89}%
  \BibitemOpen
  \bibfield  {author} {\bibinfo {author} {\bibfnamefont {I.}~\bibnamefont
  {Affleck}}, \bibinfo {author} {\bibfnamefont {D.}~\bibnamefont {Gepner}},
  \bibinfo {author} {\bibfnamefont {H.}~\bibnamefont {Schulz}}, \ and\ \bibinfo
  {author} {\bibfnamefont {T.}~\bibnamefont {Ziman}},\ }\enquote {\bibinfo
  {title} {Critical behaviour of spin-$s$ Heisenberg antiferromagnetic chains:
  analytic annd numerical results},}\ \href {\doibase
  10.1088/0305-4470/22/5/015} {\bibfield  {journal} {\bibinfo  {journal} {J.
  Phys. A}\ }\textbf {\bibinfo {volume} {22}},\ \bibinfo {pages} {511}
  (\bibinfo {year} {1989})}\BibitemShut {NoStop}%
\bibitem [{\citenamefont {Kl{\"u}mper}(1998)}]{Klu:EPJB98}%
  \BibitemOpen
  \bibfield  {author} {\bibinfo {author} {\bibfnamefont {A.}~\bibnamefont
  {Kl{\"u}mper}},\ }\enquote {\bibinfo {title} {The spin-1/2 Heisenberg chain:
  thermodynamics, quantum criticality and spin-Peierls exponents},}\ \href
  {\doibase 10.1007/s100510050491} {\bibfield  {journal} {\bibinfo  {journal}
  {Eur. Phys. J. B}\ }\textbf {\bibinfo {volume} {5}},\ \bibinfo {pages} {677}
  (\bibinfo {year} {1998})}\BibitemShut {NoStop}%
\bibitem [{Note1()}]{Note1}%
  \BibitemOpen
  \bibinfo {note} {The reader is of course free to make all quantities
  dimensionless, e.g. by dividing by $J_1$.}\BibitemShut {Stop}%
\bibitem [{\citenamefont {Hida}(1992)}]{Hid:PRB92}%
  \BibitemOpen
  \bibfield  {author} {\bibinfo {author} {\bibfnamefont {K.}~\bibnamefont
  {Hida}},\ }\enquote {\bibinfo {title} {Crossover between the Haldane-gap
  phase and the dimer phase in the spin-1/2 alternating Heisenberg chain},}\
  \href {\doibase 10.1103/PhysRevB.45.2207} {\bibfield  {journal} {\bibinfo
  {journal} {Phys. Rev. B}\ }\textbf {\bibinfo {volume} {45}},\ \bibinfo
  {pages} {2207} (\bibinfo {year} {1992})}\BibitemShut {NoStop}%
\bibitem [{\citenamefont {Regnault}\ \emph {et~al.}(1996)\citenamefont
  {Regnault}, \citenamefont {A\"in}, \citenamefont {Hennion}, \citenamefont
  {Dhalenne},\ and\ \citenamefont {Revcolevschi}}]{RAH:PRB96}%
  \BibitemOpen
  \bibfield  {author} {\bibinfo {author} {\bibfnamefont {L.~P.}\ \bibnamefont
  {Regnault}}, \bibinfo {author} {\bibfnamefont {M.}~\bibnamefont {A\"in}},
  \bibinfo {author} {\bibfnamefont {B.}~\bibnamefont {Hennion}}, \bibinfo
  {author} {\bibfnamefont {G.}~\bibnamefont {Dhalenne}}, \ and\ \bibinfo
  {author} {\bibfnamefont {A.}~\bibnamefont {Revcolevschi}},\ }\enquote
  {\bibinfo {title} {Inelastic-neutron-scattering investigation of the
  spin-Peierls system CuGe${\mathrm{O}}_{3}$},}\ \href {\doibase
  10.1103/PhysRevB.53.5579} {\bibfield  {journal} {\bibinfo  {journal} {Phys.
  Rev. B}\ }\textbf {\bibinfo {volume} {53}},\ \bibinfo {pages} {5579}
  (\bibinfo {year} {1996})}\BibitemShut {NoStop}%
\bibitem [{\citenamefont {Uhrig}\ and\ \citenamefont
  {Schulz}(1996)}]{UhS:PRB96}%
  \BibitemOpen
  \bibfield  {author} {\bibinfo {author} {\bibfnamefont {G.~S.}\ \bibnamefont
  {Uhrig}}\ and\ \bibinfo {author} {\bibfnamefont {H.~J.}\ \bibnamefont
  {Schulz}},\ }\enquote {\bibinfo {title} {Magnetic excitation spectrum of
  dimerized antiferromagnetic chains},}\ \href {\doibase
  10.1103/PhysRevB.54.R9624} {\bibfield  {journal} {\bibinfo  {journal} {Phys.
  Rev. B}\ }\textbf {\bibinfo {volume} {54}},\ \bibinfo {pages} {R9624}
  (\bibinfo {year} {1996})}\BibitemShut {NoStop}%
\bibitem [{\citenamefont {{Knetter, C.}}\ and\ \citenamefont {{Uhrig, G.
  S.}}(2000)}]{KnU:00}%
  \BibitemOpen
  \bibfield  {author} {\bibinfo {author} {\bibnamefont {{Knetter, C.}}}\ and\
  \bibinfo {author} {\bibnamefont {{Uhrig, G. S.}}},\ }\enquote {\bibinfo
  {title} {Perturbation theory by flow equations: dimerized and frustrated
  $s=1/2$ chain},}\ \href {\doibase 10.1007/s100510050026} {\bibfield
  {journal} {\bibinfo  {journal} {Eur. Phys. J. B}\ }\textbf {\bibinfo {volume}
  {13}},\ \bibinfo {pages} {209} (\bibinfo {year} {2000})}\BibitemShut
  {NoStop}%
\bibitem [{\citenamefont {Manousakis}(1991)}]{Man:RMP91}%
  \BibitemOpen
  \bibfield  {author} {\bibinfo {author} {\bibfnamefont {E.}~\bibnamefont
  {Manousakis}},\ }\enquote {\bibinfo {title} {The spin-$1/2$ Heisenberg
  antiferromagnet on a square lattice and its application to the cuprous
  oxides},}\ \href {\doibase 10.1103/RevModPhys.63.1} {\bibfield  {journal}
  {\bibinfo  {journal} {Rev. Mod. Phys.}\ }\textbf {\bibinfo {volume} {63}},\
  \bibinfo {pages} {1} (\bibinfo {year} {1991})}\BibitemShut {NoStop}%
\bibitem [{\citenamefont {Richter}(1993)}]{Ric:PRB93}%
  \BibitemOpen
  \bibfield  {author} {\bibinfo {author} {\bibfnamefont {J.}~\bibnamefont
  {Richter}},\ }\enquote {\bibinfo {title} {Zero-temperature magnetic-ordering
  in the inhomogeneously frustrated quantum heisenberg-antiferromagnet on a
  square lattice},}\ \href@noop {} {\bibfield  {journal} {\bibinfo  {journal}
  {Phys. Rev. B}\ }\textbf {\bibinfo {volume} {47}},\ \bibinfo {pages} {5794}
  (\bibinfo {year} {1993})}\BibitemShut {NoStop}%
\bibitem [{\citenamefont {Sandvik}(1997)}]{San:PRB97}%
  \BibitemOpen
  \bibfield  {author} {\bibinfo {author} {\bibfnamefont {A.~W.}\ \bibnamefont
  {Sandvik}},\ }\enquote {\bibinfo {title} {Finite-size scaling of the
  ground-state parameters of the two-dimensional Heisenberg model},}\ \href
  {\doibase 10.1103/PhysRevB.56.11678} {\bibfield  {journal} {\bibinfo
  {journal} {Phys. Rev. B}\ }\textbf {\bibinfo {volume} {56}},\ \bibinfo
  {pages} {11678} (\bibinfo {year} {1997})}\BibitemShut {NoStop}%
\bibitem [{\citenamefont {Mermin}\ and\ \citenamefont
  {Wagner}(1966)}]{MeW:PRL66}%
  \BibitemOpen
  \bibfield  {author} {\bibinfo {author} {\bibfnamefont {N.}~\bibnamefont
  {Mermin}}\ and\ \bibinfo {author} {\bibfnamefont {H.}~\bibnamefont
  {Wagner}},\ }\enquote {\bibinfo {title} {Absence of ferromagnetism or
  antiferromagnetism in one- or two-dimensional isotropic Heisenberg models},}\
  \href {\doibase 10.1103/PhysRevLett.17.1133} {\bibfield  {journal} {\bibinfo
  {journal} {Phys. Rev. Lett.}\ }\textbf {\bibinfo {volume} {17}},\ \bibinfo
  {pages} {1133} (\bibinfo {year} {1966})}\BibitemShut {NoStop}%
\bibitem [{\citenamefont {Schr{\"o}der}\ \emph {et~al.}(2010)\citenamefont
  {Schr{\"o}der}, \citenamefont {Fang}, \citenamefont {Furukawa}, \citenamefont
  {Luban}, \citenamefont {Prozorov}, \citenamefont {Borsa},\ and\ \citenamefont
  {Kumagai}}]{SFF:JPCM:10}%
  \BibitemOpen
  \bibfield  {author} {\bibinfo {author} {\bibfnamefont {C.}~\bibnamefont
  {Schr{\"o}der}}, \bibinfo {author} {\bibfnamefont {X.}~\bibnamefont {Fang}},
  \bibinfo {author} {\bibfnamefont {Y.}~\bibnamefont {Furukawa}}, \bibinfo
  {author} {\bibfnamefont {M.}~\bibnamefont {Luban}}, \bibinfo {author}
  {\bibfnamefont {R.}~\bibnamefont {Prozorov}}, \bibinfo {author}
  {\bibfnamefont {F.}~\bibnamefont {Borsa}}, \ and\ \bibinfo {author}
  {\bibfnamefont {K.}~\bibnamefont {Kumagai}},\ }\enquote {\bibinfo {title}
  {Spin freezing and slow magnetization dynamics in geometrically frustrated
  magnetic molecules with exchange disorder},}\ \href
  {http://stacks.iop.org/0953-8984/22/i=21/a=216007} {\bibfield  {journal}
  {\bibinfo  {journal} {J. Phys.: Condens. Matter}\ }\textbf {\bibinfo {volume}
  {22}},\ \bibinfo {pages} {216007} (\bibinfo {year} {2010})}\BibitemShut
  {NoStop}%
\end{thebibliography}

\end{document}